\documentclass[twocolumn,prl,floatfix,superscriptaddress,nobibnotes]{revtex4-2} 
\pdfoutput=1 
\usepackage{amsmath}
\usepackage{amsfonts}
\usepackage{amssymb}
\usepackage{physics}
\usepackage{graphicx}
\usepackage{bbm}
\usepackage[dvipsnames]{xcolor}
\usepackage{braket}
\usepackage{mathrsfs}
\usepackage{lipsum} 
\usepackage{cancel}
\usepackage{xr}
\allowdisplaybreaks

\makeatletter
\DeclareFontEncoding{LS1}{}{}
\DeclareFontSubstitution{LS1}{stix}{m}{n}
\DeclareMathAlphabet{\mathscr}{LS1}{stixscr}{m}{n}

\def \a{\hat a}
\def \c{\hat a^\dagger}

\def \omegar{\omega_r}

\def \X{\hat \sigma_x}

\def \Z{\hat \sigma_z}
\def \p{\hat \sigma_{+}}
\def \m{\hat \sigma_{-}}
\def \HE{\hat H_{E}}
\def \HEsc{\hat H_{E, s/c}}
\def \HR{\hat H_{R}}

\def \SR{\hat S_{R}}

\def \S{\hat S}

\def \tiz{\tau_{\text{i}\zeta}}
\def \tfz{\tau_{\text{f}\zeta}}

\begin{document}

\title{High-Fidelity CZ Gates in Double Quantum Dot -- Circuit QED Systems Beyond the Rotating-Wave Approximation}

\author{Guangzhao \surname{Yang}}
\affiliation{School of Physical and Mathematical Sciences, Nanyang Technological University, 21 Nanyang Link, Singapore 637371, Singapore}

\author{Marek \surname{Gluza}}
\affiliation{School of Physical and Mathematical Sciences, Nanyang Technological University, 21 Nanyang Link, Singapore 637371, Singapore}

\author{Si Yan \surname{Koh}}
\affiliation{School of Physical and Mathematical Sciences, Nanyang Technological University, 21 Nanyang Link, Singapore 637371, Singapore}

\author{Calvin Pei Yu Wong}
\affiliation{Institute of Materials Research and Engineering (IMRE), Agency for Science, Technology and Research (A*STAR), Singapore 138634, Singapore}

\author{Kuan Eng Johnson Goh}
\affiliation{School of Physical and Mathematical Sciences, Nanyang Technological University, 21 Nanyang Link, Singapore 637371, Singapore}
\affiliation{Institute of Materials Research and Engineering (IMRE), Agency for Science, Technology and Research (A*STAR), Singapore 138634, Singapore}
\affiliation{Department of Physics, Faculty of Science, National University of Singapore, Singapore 117551, Singapore}

\author{Bent Weber}
\affiliation{School of Physical and Mathematical Sciences, Nanyang Technological University, 21 Nanyang Link, Singapore 637371, Singapore}

\author{Hui Khoon Ng}
\affiliation{Yale-NUS College, Singapore 138527, Singapore}
\affiliation{Centre for Quantum Technologies, National University of Singapore, Singapore 117543, Singapore}

\author{Teck Seng \surname{Koh}}
\thanks{Corresponding author:~\href{mailto:kohteckseng@ntu.edu.sg}{kohteckseng@ntu.edu.sg}}
\affiliation{School of Physical and Mathematical Sciences, Nanyang Technological University, 21 Nanyang Link, Singapore 637371, Singapore}
\email{kohteckseng@ntu.edu.sg}

\date{\today}

\begin{abstract}
Semiconductor double quantum dot (DQD) qubits coupled via superconducting microwave resonators provide a powerful means of long-range manipulation of the qubits' spin and charge degrees of freedom. Quantum gates can be implemented by parametrically driving the qubits while their transition frequencies are detuned from the resonator frequency. Long-range two-qubit CZ gates have been proposed for the DQD spin qubit within the rotating-wave approximation (RWA). Rapid gates demand strong coupling, but RWA breaks down when coupling strengths become significant relative to system frequencies. Therefore, understanding the detrimental impact of time-dependent terms ignored by RWA is critical for high-fidelity operation. Here, we go beyond RWA to study CZ gate fidelity for both DQD spin and charge qubits. We propose a novel parametric drive on the charge qubit that produces fewer time-dependent terms and show that it outperforms its spin counterpart. We find that drive amplitude - a parameter dropped in RWA - is critical for optimizing fidelity and map out high-fidelity regimes. Our results demonstrate the necessity of going beyond RWA in understanding how long-range gates can be realized in DQD qubits, with charge qubits offering considerable advantages in high-fidelity operation.
\end{abstract}

\maketitle

Building a large-scale quantum computer requires qubits that are able to retain their quantum coherence, as well as a flexible architecture that enables both short and long range inter-qubit coupling schemes. Electrons confined to semiconductor double quantum dots (DQD)~\cite{Zwanenburg2013, Chatterjee2021, Burkard2023, Becher2023} possess both spin and charge degrees of freedom -- properties that are useful for encoding quantum information and manipulation by electric fields. Spin coherence can be improved with material enrichment to remove spinful nuclear isotopes~\cite{Tyryshkin2012, Veldhorst2014, Eng2015} or with dynamical decoupling~\cite{Alvarez2011, Muhonen2014}, while both spin and charge coherence are enhanced at optimal points in parameter space~\cite{Reed2016, Feng2021, Kratochwil2021, Pico-Cortes2021}. 
 
Inter-qubit coupling based on Heisenberg exchange~\cite{Levy2002, DiVincenzo2000} and capacitive coupling~\cite{Neyens2019, Feng2021} have limited range. Long distance coupling schemes may transfer qubit states across quantum dot arrays~\cite{Feng2018, Mills2019, Yoneda2021}, or make use of superconducting resonators in the context of circuit quantum electrodynamics (QED)~\cite{Childress2004, Burkard2006, Frey2012, Hu2012, Jin2012,  Benito2017, Tosi2017, Mi2017, Stockklauser2017,  Mi2018, Samkharadze2018, Landig2018, Scarlino2022, Warren2019, Burkard2020, Blais2021}. 

In circuit QED, photons confined in superconducting resonators serve as mediators of quantum information, as they can couple to both spin and charge degrees when resonant microwave frequencies match qubit frequencies. A single electron DQD device may be operated as a spin or charge qubit, depending on whether quantum information is encoded in the spin or charge degree of freedom. Even though direct magnetic coupling of spin to a photon is weak, by delocalizing the electron wavefunction between DQDs, a tunable electric dipole can be induced which can couple to the electric field of the photon. Both qubits take advantage of this for fast manipulation. 

Recently, ultrastrong coupling between microwave photons and a DQD charge qubit has been achieved~\cite{Scarlino2022}. Strong coupling has also been demonstrated in DQD spin and charge qubits~\cite{Mi2017, Stockklauser2017, Mi2018, Samkharadze2018} and a triple QD spin qubit~\cite{Landig2018}. Even though spin qubits have longer coherence times, the bare charge-photon coupling $g_c$ can be an order of magnitude larger than the effective spin-photon coupling $g_s$~\cite{Samkharadze2018, Scarlino2022}, allowing faster gates for the charge qubit. 


In this work, we theoretically investigate the two-qubit Controlled-Z (CZ) gate between a pair of DQD qubits coupled via microwave photons in a superconducting resonator. Detuning the qubit transition frequencies from the resonator frequency avoids entangling the qubits with the microwave photons. In this regime, quantum gates can be implemented by parametrically driving the qubits, giving rise to a sideband transition. 

Existing works~\cite{Abadillo-Uriel2021, Srinivasa2016, Srinivasa2023} studied the CZ gate for spin qubits in the rotating-wave approximation (RWA), while charge qubits have received less attention. However, RWA breaks down when coupling strengths become significant relative to system frequencies. Given recent progress towards ultrastrong coupling, understanding the detrimental impact of time-dependent terms ignored by RWA is critical for high-fidelity operation.  We go beyond RWA to show how CZ gates may be optimized for both DQD spin and charge qubits. 

\begin{figure}
    \includegraphics[width=0.95\linewidth]{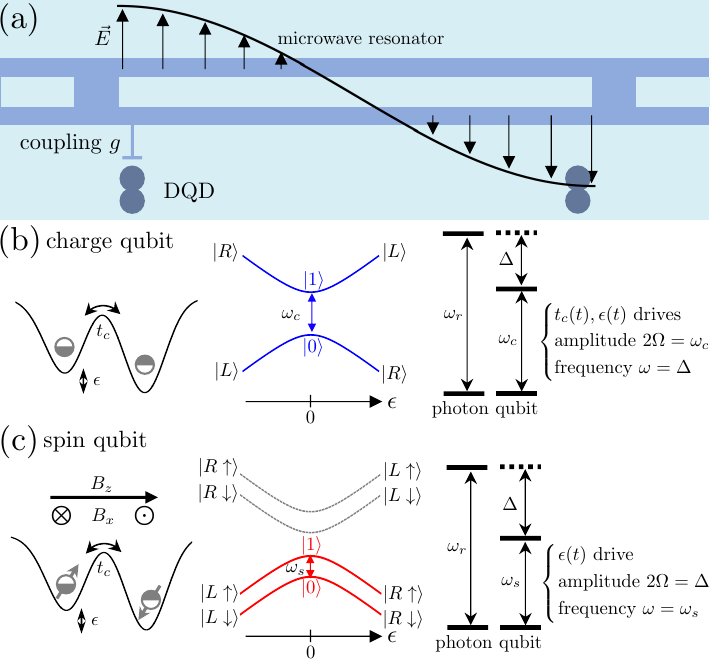}
    \caption{\label{fig:setup} (a) Schematic of two spatially separated DQD qubits in a circuit QED system. (b) Single-electron DQD charge qubit and energy scales. Tunnel coupling $t_c$ and orbital detuning $\epsilon$ control the charge qubit frequency $\omega_c$. Logical states (blue) are eigenstates of $\hat H_\text{DQD}$. Simultaneous  $t_c (t)$ and $\epsilon(t)$ drives with amplitude $2\Omega = \omega_c$ and drive frequency $\omega = \Delta$ enables red-sideband transitions. (c) Single-electron DQD spin qubit. The Zeeman $B_z$ field spin-splits each orbital; a spin-orbit field $B_x$ admixes spin and orbital states, enabling electrical control of the logical states (red). A resonant $\epsilon(t)$ drive with amplitude $2\Omega = \Delta$ enables red-sideband transitions. }\label{fig:F1}
\end{figure}


We consider a single-electron DQD, which may be operated in the charge qubit or spin qubit regime. In the CZ gate sequence, when one qubit is coupled to the resonator, the other is uncoupled. Therefore, for simplicity, we only include one qubit in the expressions for the Hamiltonians. Subscripts $r, c, s$ denote resonator, charge and spin. The Hamiltonians for a charge and spin qubit coupled to a resonator are 
\begin{align}
	\hat H_{c} &=  \omega_r \c \a + \hat H_\text{DQD} + \hat H_\text{int}, \\
	\hat H_{s} &=  \omega_r \c \a + \hat H_\text{DQD} + \hat H_\text{B} + \hat H_\text{int}.
\end{align}
Here, $\omegar$ is the resonator frequency and $\hat H_\text{DQD} = (\epsilon \hat\tau_z + t_c \hat\tau_x )/2$ comprises orbital detuning $\epsilon$ and tunnel coupling $t_c$. $\hat H_\text{int} = g_c \hat \tau_z \left(\a + \c \right)$ is the bare coupling between the charge dipole and resonator, where $\hat \tau_z \equiv \ket{L}\bra{L} - \ket{R}\bra{R}$ is given in the DQD orbital basis. Together, these describe the charge qubit. For the spin qubit, the additional magnetic term $\hat H_\text{B} = (B_z \hat S_z + B_x \hat \tau_z \hat S_x)/2$ completes the description. It comprises a Zeeman field, and a spin-orbit interaction term which may be intrinsic~\cite{Golovach2006, Weber2018} or engineered~\cite{Tosi2017, Hsueh2024}, e.g. with a micromagnet~\cite{Xian2014, Mi2017, Benito2017}. Here, $\hat S_z \equiv \ket{\uparrow}\bra{\uparrow} - \ket{\downarrow}\bra{\downarrow}$. At the sweet spot $\epsilon = 0$, coherence and coupling are enhanced simultaneously.

Following Ref.~\cite{Srinivasa2023}, a perturbative drive on orbital detuning $\epsilon(t) = - \epsilon_0 \cos(\omega t + \phi)$ at the sweet spot produces a transverse drive in the rotating frame defined by $e^{i (\pi/4) \hat \tau_y}$. When $t_c > B_z$, projection onto the ground orbital gives the effective spin-resonator Hamiltonian,
\begin{align}\label{eq:transverse coupling}
	\hat H_{s}^\prime = &\omegar \c\a + \frac{\omega_s}{2} \Z -  2 \Omega \cos (\omega t + \phi) \X  + g_s (\a + \c) \X. 
\end{align}
Here, $\Z \equiv \ket{1}\bra{1} - \ket{0}\bra{0}$, spin splitting $\omega_s \approx B_z$, spin-photon coupling $g_s \equiv g_c |\sin (\Phi/2)|$ and effective drive amplitude $2 \Omega \equiv \epsilon_0 |\sin (\Phi/2)|$, with $\Phi \equiv \arctan \tfrac{B_x}{t_c-B_z} \in (-\pi/2,\pi/2)$.

For the charge qubit, since tunnel coupling is non-negative, we drive $t_c(t) =  2\Omega \sin(\frac{\omega t + \phi}{2})^2$ and define $2\Omega$ as drive amplitude, for ease of qubit comparison. As such, orbital detuning needs to be driven as $\epsilon(t) = 2\Omega \sqrt{1 - \sin(\frac{\omega t + \phi}{2})^4}$, to maintain constant qubit frequency $\omega_c \equiv \sqrt{\epsilon^2 + t_c^2} =  2\Omega$. 
The $t_c(t)$ drive contains a d.c. component and one Fourier frequency $\omega$, but the $\epsilon(t)$ drive contains infinite Fourier components, demanding a high bandwidth waveform generator in practice. 

In the eigenbasis of $\hat H_\text{DQD}$, which are the logical states we denote similarly with $\ket{1}, \ket{0}$, the drives lead to a time-dependent interaction of the charge qubit with the resonator,
\begin{align} \label{eq: charge Hamiltonian}
    \hat H_{c}^\prime = \omegar \c\a + \frac{\omega_c}{2} \Z + \left( g_z(t) \Z - g_x(t) \X \right)(\a + \c), 
\end{align}
where $g_x(t) = g_c t_c(t)/2\Omega$ and $g_z(t) = g_c \epsilon(t)/2\Omega$. 


Sidebands are generated under resonant conditions with amplitude modulation. The spin qubit is driven on-resonance $\omega = \omega_{s}$, with drive amplitude equal to the qubit-resonator detuning $2\Omega = \Delta \equiv \omega_r - \omega_{s}$. Interestingly, the charge qubit $t_c(t)$ drive is resonant with the qubit-resonator detuning $\omega = \Delta \equiv \omega_r - \omega_{c}$, while its drive amplitude is equal to the qubit frequency $2\Omega = \omega_c$. Both parametric drives produce red-sidebands, which can be seen by transforming Eqs.~\eqref{eq:transverse coupling} and~\eqref{eq: charge Hamiltonian} to their respective rotating frames (see Sec.II in ~\cite{SI}). After RWA, they yield the same red-sideband Hamiltonian, given by 
\begin{equation}\label{eq: RWA Hamiltonian}
	\HR = \frac{ig}{2} (e^{i \phi}\a\p - e^{- i \phi} \c\m),
\end{equation}
where $g = g_s$ for the spin qubit and $g = g_c/2$ for the charge qubit. 

The red-sideband is time-independent by design: drive parameters were chosen to cancel the oscillatory behaviour of the terms of Eq.~\eqref{eq: RWA Hamiltonian}, which describe exchanges of excitations between qubit and resonator. Residual oscillatory terms are fast and ignored within RWA. Consequently, all dependence on drive amplitude and accordingly, qubit-resonator detuning, vanishes in the approximate dynamics governed by Eq.~\eqref{eq: RWA Hamiltonian}. As such, any limitations on the tunability of parameters and optimal gate fidelity cannot be derived with RWA. Therefore, we analyze the exact Hamiltonians without RWA, which are 
\begin{align}
	\hat H_{E,s}(t) &= \HR + \hat V_\text{red}(t) + \hat V_\text{blue}(t) + \hat V_\text{long}(t) + \hat V_\text{qubit}(t), \label{eq:exact spin hamiltonian} \\
    \hat H_{E,c}(t) &= \HR + \hat V_{\text{red}, t_c}(t) + \hat V_{\text{blue}, t_c} (t) + \hat V_{\text{long}, \epsilon} (t). \label{eq:exact charge hamiltonian}
\end{align}
Explicit expressions are given in Sec.II of~\cite{SI}.


All time-dependent terms contain oscillatory functions. $\hat V_\text{red}(t)$ and $\hat V_{\text{red}, t_c}(t)$ contain $\a \p, \c \m$ terms that generate fast rotations to the red-sideband directly. Counter-rotating terms of the blue sideband $\hat V_\text{blue}(t)$ and $\hat V_{\text{blue}, t_c}(t)$ induce doubly excited transitions of the form $\c \hat{\sigma}_{+}, \a \hat{\sigma}_{-}$. Longitudinal coupling terms $\hat V_\text{long}(t)$ and $\hat V_{\text{long}, \epsilon}(t)$ generate rotations in $\a \Z$ and $\c \Z$. The oscillatory red and blue-sideband and longitudinal terms have fast Fourier components $\varpi \in \pm \{2\Omega, 4\Omega, 2(\omega - \Omega), 2(\omega + \Omega), 2(\omega + 2\Omega), 2\omega\}$, with magnitudes of order $g$.  The usual RWA argument applies: when $|g / \varpi | \ll 1$, the deleterious contribution to the dynamics is small. As such, the appearance of $\Omega$ in the Fourier component leads to a lower bound on drive amplitude, $2\Omega \gg g$. 

Significantly, the $\hat V_\text{qubit}(t)$ term, which causes spurious qubit rotations $\hat{\sigma}_{\pm,z}$, is absent from the exact charge qubit Hamiltonian compared to its spin counterpart. This term represents deviations from ideal parametric drive, and is of magnitude $\Omega$ with Fourier components $\varpi \in \pm  \{2\omega, 2(\omega - \Omega), 2(\omega + \Omega) \}$. The RWA requirement $|\Omega / \varpi | \ll 1$ leads to an upper limit on drive amplitude, beyond which the spin qubit becomes vulnerable to time-dependent errors. It suggests that the spin qubit has an optimal drive amplitude between this upper bound and the lower bound from the $\hat V_\text{red}(t)$ term. 

Arising from the $\epsilon(t)$ drive on the charge qubit, oscillatory terms of $\hat V_{\text{long}, \epsilon}(t)$ become constant when $2\Omega = \omega_r (1 - 1/p)$, where $p$ are non-zero integers, suggesting the failure of RWA at these points. Fortunately, the impact on fidelity is lessened by the diminishing magnitude of progressively larger frequency terms. (See Sec.IV of~\cite{SI}.)

The most significant impact on fidelity for the charge qubit occurs at $2\Omega = \omega_r$, when the $\hat V_{\text{red},t_c}(t)$ term becomes time-independent. This is the case when qubit frequency, and consequently the drive frequency, approaches zero. Therefore, the maximum drive amplitude for the charge qubit is inherently restricted. 


Next, we discuss the CZ gate which is implemented by a five-stage sequence of the RWA red-sideband, reported in Ref.~\cite{Abadillo-Uriel2021}. The CZ gate works in the zero-photon subspace, and is 
\begin{align}
	\hat U_\text{CZ} = &\hat M_0 \hat Z^{(1)} \left( \frac{\pi}{\sqrt{2}} \right) \hat Z^{(2)} \left(- \frac{\pi}{\sqrt{2}}\right) \SR^{(2)} \left(\pi, \pi \right) \SR^{(1)} \left(\frac{\pi}{2}, 0 \right)  \nonumber \\
				& \times \SR^{(1)} \left(\sqrt{2} \pi, \frac{\pi}{2} \right) \SR^{(1)} \left(\frac{\pi}{2}, \pi \right) \SR^{(2)} \left(\pi, 0 \right), \label{eq: Ucz}
\end{align}
where superscript index $(j)$ denotes the coupled qubit, and the red-sideband evolution is $\SR^{(j)}(g \Delta t, \phi) = e^{- i \Delta t \HR^{(j)}(g,\phi)}$. The exact evolution $\hat U_{E, c/s}$ is calculated similarly for each qubit system, replacing $\SR^{(j)}$ with $\S_{E,c/s}^{(j)}$, where $\S_{E,c/s}^{(j)} = \mathcal{T}_{\leftarrow} \exp[-i \int d t \, \HEsc^{(j)}(g,\phi)]$. These are followed by single-qubit $z$-rotations $\hat{Z}^{(j)}(\theta) = e^{-i \theta\hat{\sigma}_z^{(j)}/2 }$ and zero-photon projection  $\hat{M}_0 \equiv \ket{0}_r\!\bra{0}$.   

The RWA conditions in the preceding discussion translate naturally to conditions on the quantum gate. Gate fidelity~\cite{Wang2008} between the ideal and exact evolution is $F = \frac{1}{d} \abs{\Tr (U_\text{CZ}^{\dag} U_{E} )}$, where $d=4$. With this, we map out regions of high fidelity in dimensionless parameter space, rescaled by $g$. Henceforth, we work in dimensionless frequencies $\tilde\omega_r \equiv \omegar/g$, $\tilde\omega_{c/s} \equiv \omega_{c/s}/g$, $\tilde\Omega \equiv \Omega/g$ and dimensionless time $\Delta \tau \equiv  g \Delta t$. 

\begin{figure}
    \centering
    \includegraphics[width=0.95\linewidth]{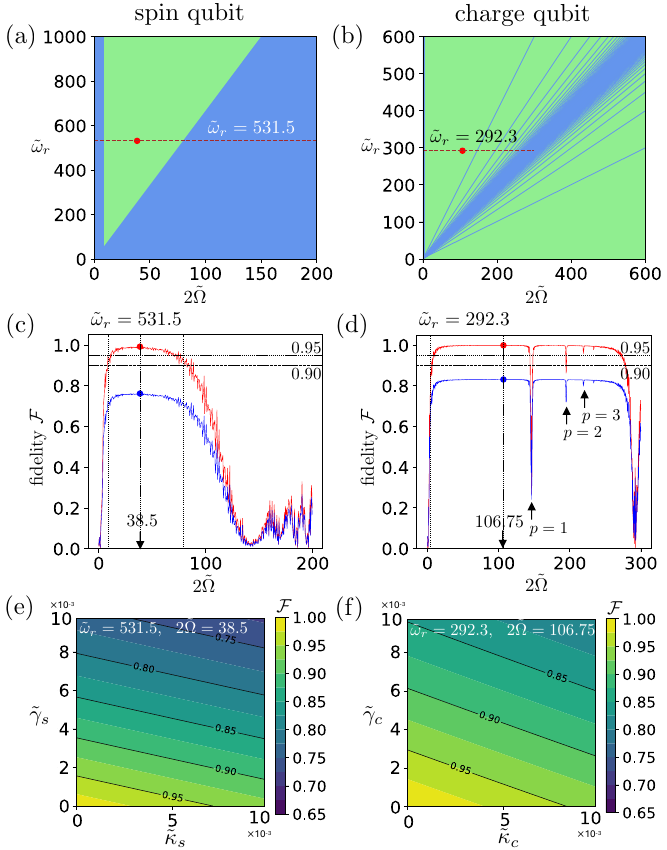}
    \caption{Gate fidelity of spin (left column) and charge (right column) qubits. (a, b) Map of high (green) and low (blue) fidelity regions against rescaled resonator frequency and drive amplitudes. (b, c) Fidelity linecuts at $\tilde\omega_r = 531.5$ and $292.3$, against rescaled drive amplitude, with (blue) and without (red) decoherence, for rates $\tilde\kappa_s = \tilde\kappa_c = \tilde\gamma_s = \tilde\gamma_c = 0.008$. Highest fidelities for spin and charge qubits are $99.199\%$ and $99.949\%$ (red circles), which drop to  $76\%$ and $83\%$ (blue circles) with decoherence. (e, f) Fidelity against rescaled photon loss and dephasing rates at optimal drive amplitudes and resonator frequencies used in panels (a--d).}
    \label{fig:F2}
\end{figure}

An approximation to fidelity is calculated using the Neumann series expansion of the unitaries to second order (see Sec.V in~\cite{SI}). By imposing strict numerical thresholds on RWA conditions, we obtain expressions for the boundaries defining regions of high fidelity: 
\begin{align}
    \tilde\Omega &= \sqrt{2}\pi, \tilde\Omega = \frac{\tilde\omega_r}{\varrho} ~ \text{(spin qubit)}, \label{eq: spin-qubit high-F condition}\\
         \tilde\Omega &= \frac{\sqrt{2}\pi}{2}, \tilde\Omega = \frac{\tilde\omega_r}{2}, \tilde\Omega = \frac{\tilde\omega_r}{2}\left(1 - \frac{1}{p}\right) ~ \text{(charge qubit)}, \label{eq: charge-qubit high-F condition}
\end{align}
where $\varrho \approx 13.347$. These are shown in Fig.~\ref{fig:F2}(a,b). The spin qubit has a single, triangular high-fidelity region. In contrast, the charge qubit has multiple high-fidelity regions which are symmetrical about the diagonal $2\tilde \Omega = \tilde \omega_r$. These high-fidelity slices become sharper and narrower around the diagonal, which appears to diminish the practicality of operating in those regions. Fortunately, as mentioned in the preceding discussion on the breakdown of RWA from the $\epsilon(t)$ drive, at the boundaries of these narrow slices, the loss in fidelity is very small. In addition, for the same $\tilde \omega_r$, the high-fidelity regions for the charge qubit are broader than that of the spin qubit, giving it greater flexibility in operation. 

We model decoherence of the quantum gate by computing the Lindblad superoperator,  $\mathscr{L} \, \cdot = -i [ \HE, \cdot] + \sum_{\ell} \tilde\gamma_\ell \left(\hat{L}_\ell \cdot \hat{L}_\ell^\dag - \tfrac{1}{2} \{\hat{L}_\ell^\dag \hat{L}_\ell, \cdot\} \right)$, where the index $\ell = p, q^{(1)}, q^{(2)}$ represent photon loss and dephasing of qubits 1, 2.  In the Schr\"{o}dinger picture, their corresponding Lindblad operators are $\a$ and $\Z^{(1)}, \Z^{(2)}$. Time evolution is given by the semigroup superoperator $\mathscr{S}(\zeta, \tfz, \tiz) = \mathcal{T}_{\leftarrow} \exp \left[ \int_{\tiz} ^{\tfz} d\tau \mathscr{L}(\zeta, \tau) \right]$ where qubit-resonator interactions are labeled by $\zeta$. The exact quantum gate with decoherence is 
\begin{align} \label{eq: semigroup evolution CZ gate}
    \mathscr{S}_{E} = \mathscr{M}_0 \mathscr{Z} \left( \mathcal T_{\leftarrow}\prod_{\zeta=1}^{5} \mathscr{S}(\zeta; \tfz, \tiz) \right),
\end{align}
where $\mathscr{M}_0$ and $\mathscr{Z}$ are projection and single-qubit superoperators. Gate fidelity with decoherence is thus~\cite{poulin2011quantum, greenbaum2015introduction}
\begin{equation}
	\mathcal{F} = \frac{1}{d^2}\left| \text{Tr}\left( \mathscr{U}_\text{CZ}^\dag \mathscr{S}_{E} \right)\right|,
\end{equation}
where $\mathscr{U}_\text{CZ}$ is the superoperator corresponding to $\hat U_\text{CZ}$. Single-qubit rotations are very fast (sub-ns) compared to qubit-resonator gate times (sub-$\mathrm{\mu}$s). Therefore, we assume all decoherence effects arise during the qubit-photon interaction stages. 


Taking experimental parameters from Ref.~\cite{Mi2018}, we compute fidelities using $\omega_r / 2\pi = 5.846 ~\mathrm{GHz}$, $g_s/2\pi = 11~\mathrm{MHz}$ and $g_c/2\pi =  40~\mathrm{MHz}$, corresponding to $\tilde \omega_r = 531.5$ and $292.3$ for spin and charge qubits. 

In Fig.~\ref{fig:F2}(c,d), gate fidelities are plotted against rescaled drive amplitude $2\tilde \Omega$. The plots verify the analyses of the exact Hamiltonians: both qubits have fidelities greater than $99\%$ when RWA is satisfied. The charge qubit has wider optimal regions with flatter fidelity curves, demonstrating greater robustness across drive amplitudes. The drop in fidelity is greatest when $2\tilde \Omega = \tilde \omega_r$, and is successively smaller for larger $p$. At optimal drive amplitudes, fidelities of $99.199\%$ and $99.949\%$ can be achieved for the spin and charge qubits. 

\begin{figure}
    \includegraphics[width=0.95\linewidth]{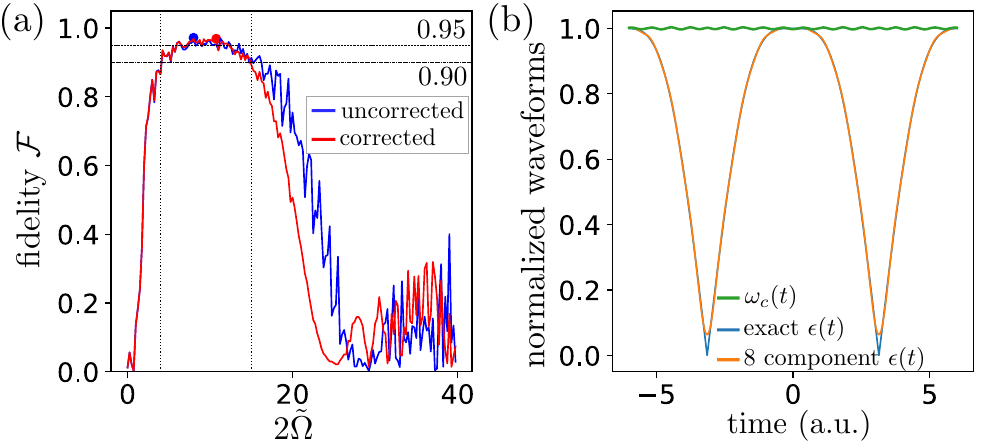}
    \caption{(a)  Spin qubit fidelity for $\tilde\omega_r = 158$, with (red) and without (blue) gate time correction~\cite{SI}.  The curve is smoothed but the best fidelity (circles) is not improved.  (b)  Exact $\epsilon(t)$ waveform (blue) compared against realistic waveform (orange) composed with the first 8 Fourier components. The latter produces a good approximation of constant qubit frequency $\omega_c(t)$ (green). }
\label{fig:F3}
\end{figure}

Next, we attempt to improve spin qubit fidelity by accounting for the Bloch-Siegert shift~\cite{Ashhab2007, lu2012effects}. This leads to a gate time correction arising from eliminating residual $\Z$ terms from strong driving. It is applicable only to the spin qubit due to the presence of $\hat V_\text{qubit}(t)$. In the time-dependent rotating frame defined by $\exp\left[-\tfrac{i\Omega}{2\omega}\sin(2\omega t + 2\phi)\hat\sigma_z \right]$, gate time correction is calculated using the Jacobi-Anger identity (see Sec.VIII in~\cite{SI}) and fidelity replotted in Fig.~\ref{fig:F3}(a). We observe that this approach only smoothens the fast oscillations in fidelity; the time-dependence of $\Z$ gets transferred to other qubit terms in $\hat V_\text{qubit}(t)$ and thus, does not improve fidelity.

With decoherence, the fidelity curves in Fig.~\ref{fig:F2}(c,d) are similar, but best fidelities drop to $76\%$ and $83\%$ for the spin and charge qubits. Our calculations used decoherence rates $\tilde\kappa_s = \tilde\kappa_c = \tilde\gamma_s = \tilde\gamma_c = 0.008$. Plotting the fidelity at the optimal drive amplitude against decoherence rates in Fig.~\ref{fig:F2}(e,f), we observe that both qubits are more sensitive to dephasing than photon loss. From these results, fidelities greater than $95\%$ require
\begin{align}
	\frac{\tilde\gamma_s}{1.554} + \frac{\tilde\kappa_s}{7.150} & \leq 10^{-3} ~\text{(spin qubit)}, \label{eq:ultrastrong coupling spin qubit}\\
	\quad \frac{\tilde\gamma_c}{2.927} + \frac{\tilde\kappa_c}{8.250} & \leq 10^{-3}~\text{(charge qubit)}. \label{eq:ultrastrong coupling charge qubit}
\end{align}
Notably, Ref.~\cite{Scarlino2022} reported rates that give $\sim 8\times 10^{-3}$ for the left-hand-side of inequality~\eqref{eq:ultrastrong coupling charge qubit}. In contrast, Refs.~\cite{Mi2017, Stockklauser2017, Mi2018, Samkharadze2018} reported rates $1-3$ orders of magnitude larger than required by inequalities~\eqref{eq:ultrastrong coupling spin qubit}, \eqref{eq:ultrastrong coupling charge qubit}.

Fig.~\ref{fig:F2}(e,f) also show that the charge qubit is more robust against decoherence, due in part to the stronger qubit-photon coupling and hence faster gates. The other reason is that in the ideal case, the charge qubit does not suffer from spurious effects of the parametric drive, compared to the spin qubit. However, in reality, there will be deviations from the ideal waveform due to finite bandwidths. For example, a state-of-the-art waveform generator of 50~GHz bandwidth can only compose 2 Fourier components of the $\epsilon(t)$ drive at the optimal drive amplitude shown in Fig.~\ref{fig:F2}(d). This will lead to large waveform deviations and lowered fidelity, but it can be mitigated by driving at a lower frequency. This implies a higher qubit frequency, which is dynamically tunable, and a larger drive amplitude for the same resonator frequency. Fortunately, the broad high-fidelity region of the charge qubit offers considerable flexibility in the choice of operating point. For instance, moving to a larger drive amplitude and qubit frequency $2\tilde\Omega = \tilde \omega_c =247$ gives a smaller drive frequency $\tilde \omega = 45.3$, which allows 8 Fourier components to be composed for the $\epsilon(t)$ drive. This gives a residual deviation of less than 0.79\% per cycle and provides a good approximation of constant qubit frequency. Deviations are tiny and high frequency oscillations produce small effects on gate times that can be corrected by a similar time-dependent transformation as discussed for the spin qubit. We find a reduction in fidelity to $99.291\%$ arising mainly from moving away from the optimal drive amplitude; the contribution to infidelity from the finite bandwidth is marginal. This remains slightly higher than optimal spin qubit fidelity. Therefore, the charge qubit is not limited by bandwidth constraints.

Finally, the CZ gate requires an initial, disentangled zero-photon state. It evolves  out of the zero-photon subspace, entangles with the qubits, before finally returning to the disentangled zero-photon state. Importantly, a non-zero initial photon population generates an evolution that is not a CZ gate, and produces qubit-photon entanglement at the end of the sequence. Therefore, any initial finite photon population leads to gate infidelity and two-qubit mixed states. Hence, even though a driven resonator can also induce sidebands~\cite{Srinivasa2016}, such a drive generates finite photon population, rendering this option unfavorable.

In conclusion, we studied the exact CZ gate fidelity for DQD spin and charge qubits in a circuit QED architecture and identified regimes for optimal fidelity. We proposed a novel parametric drive for the charge qubit with fewer undesired time-dependent terms. Although the cost is higher bandwidth requirements, it is not a limiting constraint. Weighed against experimental overheads of the spin qubit, e.g. micromagnet fabrication, we believe that the charge qubit is simpler to fabricate and offers considerable advantages in terms of fidelity, range of optimal drive parameters and robustness against decoherence. Our results demonstrate the necessity of going beyond RWA  in understanding how long-range gates can be implemented in DQD qubits, and communicate a renewed importance of the DQD charge qubit in realizing high-fidelity gate operations.


{\em Acknowledgements. --} This project was supported by a Singapore National Research Foundation Quantum Engineering Programme 2.0 grant (NRF2021-QEP2-02-P07) and a Singapore Ministry of Education Tier 2 Academic Research Fund (T2EP50222-0038).


\bibliographystyle{apsrev4-2}
\bibliography{GZ_Yang_References}

\end{document}



\title{Supplemental Material for: High-Fidelity CZ Gates in Double Quantum Dot -- Circuit QED Systems Beyond the Rotating-Wave Approximation}

\author{Guangzhao \surname{Yang}}
\affiliation{School of Physical and Mathematical Sciences, Nanyang Technological University, 21 Nanyang Link, Singapore 637371, Singapore}

\author{Marek \surname{Gluza}}
\affiliation{School of Physical and Mathematical Sciences, Nanyang Technological University, 21 Nanyang Link, Singapore 637371, Singapore}

\author{Si Yan \surname{Koh}}
\affiliation{School of Physical and Mathematical Sciences, Nanyang Technological University, 21 Nanyang Link, Singapore 637371, Singapore}

\author{Calvin Pei Yu Wong}
\affiliation{Institute of Materials Research and Engineering (IMRE), Agency for Science, Technology and Research (A*STAR), Singapore 138634, Singapore}

\author{Kuan Eng Johnson Goh}
\affiliation{School of Physical and Mathematical Sciences, Nanyang Technological University, 21 Nanyang Link, Singapore 637371, Singapore}
\affiliation{Institute of Materials Research and Engineering (IMRE), Agency for Science, Technology and Research (A*STAR), Singapore 138634, Singapore}
\affiliation{Department of Physics, Faculty of Science, National University of Singapore, Singapore 117551, Singapore}

\author{Bent Weber}
\affiliation{School of Physical and Mathematical Sciences, Nanyang Technological University, 21 Nanyang Link, Singapore 637371, Singapore}

\author{Hui Khoon Ng}
\affiliation{Yale-NUS College, Singapore 138527, Singapore}
\affiliation{Centre for Quantum Technologies, National University of Singapore, Singapore 117543, Singapore}

\author{Teck Seng \surname{Koh}}
\affiliation{School of Physical and Mathematical Sciences, Nanyang Technological University, 21 Nanyang Link, Singapore 637371, Singapore}

\maketitle

\def \id{\mathbbm{\hat 1}}
\def \Ho{\hat H_r}
\def \Hq{\hat H_{\text{qubit}}}
\def \a{\hat a}
\def \c{\hat a^\dagger}
\def \Hint{\hat H_{\text{int}}}
\def \Hrot{\hat H_{\text{rot}}}
\def \Hoq{\hat H_{oq}}
\def \D{\hat D(\alpha(t))}
\def \Urot{\hat U(t)}
\def \omegao{\omega_r}
\def \omegaq{\omega_q}
\def \omegad{\omega_d}
\def \I{\hat \sigma_0}
\def \X{\hat \sigma_x}
\def \Y{\hat \sigma_y}
\def \Z{\hat \sigma_z}
\def \p{\hat \sigma_{+}}
\def \m{\hat \sigma_{-}}
\def \HE{\hat H_E}
\def \HR{\hat H_R}

\def \Ij{\hat \sigma^{(j)}_0}
\def \Xj{\hat \sigma^{(j)}_x}
\def \Yj{\hat \sigma^{(j)}_y}
\def \Zj{\hat \sigma^{(j)}_z}
\def \pj{\hat \sigma_{+}^{(j)}}
\def \mj{\hat \sigma_{-}^{(j)}}

\def \SR{\hat S_R}
\def \SE{\hat S_E}
\def \S{\hat S}

\def \ti{\tau_{\text{i}}}
\def \tf{\tau_{\text{f}}}
\def \tiz{\tau_{\text{i}\zeta}}
\def \tfz{\tau_{\text{f}\zeta}}

\def \opvec{\hat \Theta}

\def \J{\hat J}
\def \JRz{\hat J_{Rs}}
\def \JEz{\hat J_{Es}}
\def \JRy{\hat J_{Rs'}}
\def \JEy{\hat J_{Es'}}

\def \Cp{\hat C^{+}}
\def \C{\hat C}
\def \A{\hat A}
\def \Ad{\hat A^\dag}

\tableofcontents
\clearpage

\section{Charge qubit parametric drive}
The charge qubit Hamiltonian is $\hat H_\text{DQD} = \frac{1}{2} (\epsilon \hat \tau_z + t_c \hat \tau_x )$. In the eigenbasis, $\hat H^\prime_\text{DQD} = \hat U^\dag \hat H_\text{DQD} \hat U = \frac{\omega_c}{2} \Z$, where $\omega_c = \sqrt{\epsilon^2 + t_c^2}$ and $\hat U = \cos\frac{\theta}{2} \hat I - i \sin\frac{\theta}{2} \hat \tau_y$ and $\tan \theta = t_c/\epsilon$. 
The dipole-field interaction is $\hat H_\text{int} = g_c \hat \tau_z \left(\a + \c \right)$. In the eigenbasis of $\hat H_\text{DQD}$, it becomes
\begin{align}
	\hat H_\text{int}^\prime = \left(\a + \c\right)\left(g_z \Z - g_x \X \right),
\end{align}
where $g_x = g_c \sin \theta = g_c \frac{t_c}{\omega_c}$, and $g_z = g_c \cos\theta = g_c \frac{\epsilon}{\omega_c}$. Since tunnel coupling is non-negative, a drive  applied to $t_c$ can be written as, 
\begin{align}
	t_c(t) = t_{c0} + \Omega  \sin(\omega t + \phi^\prime) \ge 0, 
\end{align}
where $t_{c0}$ is a positive constant and $\Omega $ is the amplitude. The condition $t_c \ge 0$ leads to $t_{c0} \geq |2\Omega |$. Orbital detuning is then 
\begin{align}
	\epsilon(t) =\sqrt{ \omega_c^2 - (t_{c0} + \Omega  \sin(\omega t + \phi^\prime))^2}.
\end{align}
Since $\epsilon(t)$ has to be real, this implies $t_{c0} + |\Omega  | \leq \omega_c$.

The red-sideband transition terms arise from a unitary transformation of the $g_x \hat \sigma_x$ term. To maximize coupling strength, and satisfy the preceding conditions simultaneously, we choose $\Omega  = t_{c0} = \omega_c/2$. Therefore,  
\begin{align}
	g_x (t) =&  g_c\frac{ t_c(t)}{\omega_c} = g_c\frac{\Omega }{\omega_c}  (1 + \sin(\omega t + \phi^\prime)) =  \frac{g_c}{2}  (1 - \cos(\omega t + \phi^\prime + \pi/2)) \\
	=& g_c \sin (\frac{\omega t + \phi}{2})^2,
\end{align}
where we have made the substitution $\phi = \phi^\prime + \pi/2$ in the last line. Since $g_z(t)^2 + g_x(t)^2= g_c^2$, we have the qubit-field interaction,
\begin{align}
	\hat H_\text{int}^\prime = g_c \left(\a + \c\right)\left(\sqrt{1- \sin (\frac{\omega t + \phi}{2})^4 } \Z - \sin (\frac{\omega t + \phi}{2})^2 \X \right),
\end{align}
from the simultaneous parametric drives
\begin{align}
	t_c(t) =&\Omega (1 +  \sin(\omega t + \phi^\prime)) = 2\Omega  \sin (\frac{\omega t + \phi}{2})^2,\\
	\epsilon(t) =& \sqrt{ \omega_c^2 - \Omega^2 (1+\sin(\omega t + \phi^\prime))^2} = 2\Omega \sqrt{ 1 -  \sin (\frac{\omega t + \phi}{2})^4}.
\end{align}

\section{Derivation of exact Hamiltonians}
To derive the exact and red-sideband Hamiltonians from the spin and charge qubit Hamiltonians $\hat H_s^\prime, \hat H_c^\prime$ in the main text, we use the unitary transformations
\begin{align} 
    \hat U_s(t) &= e^{-i \omega t \left( \c\a + \sum_q \frac{1}{2}\Z^{(q)}\right)} e^{-i \left(\phi + \frac{\pi}{2}\right) \frac{1}{2}\sum_q \Z^{(q)}} e^{ -i t \left( \Delta_0 \c\a + \Omega^{(j)} \Y^{(j)} \right)} e^{i \frac{\pi}{4} \sum_q \X^{(q)}} \quad \text{(spin)}, \label{eq: spin unitary transformation}\\
    \hat U_c(t) &= e^{-\sum_q \frac{i\pi}{4} \Z^{(q)}} e^{-\sum_q \frac{i\pi}{4} \Y^{(q)}} e^{-\sum_q \frac{i\pi}{4}\Z^{(q)}} e^{ -i \omega_r t \c\a - i \sum_q \left(\Omega t - \frac{\pi}{4}\right) \Y^{(q)}} e^{i \frac{\pi}{4} \sum_q \X^{(q)}}  \quad \text{(charge)}, \label{eq: charge unitary transformation}
\end{align}
where the index $q =1,2$, labels the qubits. The unitary transformations allow us to obtain the exact Hamiltonians, which are functions of the coupling strength, drive parameters $(\Omega, \omega, \phi)$ and time:
\begin{align}
    \hat H_{E, s}^{(j)} (g_s, \Omega, \omega, \phi, t) &= \HR^{(j)} + \hat V_\text{red}^{(j)}(t) + \hat V_\text{blue}^{(j)}(t) + \hat V_\text{long}^{(j)}(t) + \hat V_\text{qubit}^{(j)}(t) \label{eq: spin exact Hamiltonian},\\
    \hat H_{E, c}^{(j)} (g_c / 2, \Omega, \omega, \phi, t) &= \HR^{(j)} + \hat V_{\text{red}, t_c}^{(j)}(t) + \hat V_{\text{blue}, t_c}^{(j)} (t) + \hat V_{\text{long}, \epsilon}^{(j)}(t)  \label{eq: charge exact Hamiltonian}.
\end{align}
Here, we explicitly indicate with a superscript index $j$ to label the qubit that is coupled to the photon. This was omitted for notational simplicity in the main text. 

After applying RWA, both the exact Hamiltonians simplify to the red-sideband Hamiltonian, which is now only a function of effective qubit-photon coupling strength $g$ and driving phase $\phi$,
\begin{align}  \label{eq: RWA Hamiltonian}
    \HR^{(j)} (g \in \{g_s, g_c/2\}, \phi) = \frac{ig}{2} (e^{i \phi}\a\pj - e^{- i \phi} \c\mj), 
\end{align}
We note that in the red-sideband Hamiltonian, the effective charge qubit-photon interaction strength is $g_c/2$ compared to the effective spin qubit-photon interaction strength $g_s$.

The time-dependent terms in spin qubit Hamiltonian are
\begin{align}
    \hat V_\text{qubit}^{(j)}(t) &= \Omega e^{2i\Omega t} \sin(2\omega t + 2\phi) \pj + \Omega e^{-2i\Omega t} \sin(2\omega t + 2\phi) \mj + \Omega \cos(2\omega t + 2\phi) \Zj\\
    \hat V_\text{red}^{(j)}(t) &= \frac{ig_s}{2}\left(-e^{-i2\omega t - i\phi} \a \pj + e^{i2\omega t + i\phi} \c \mj \right) \\
    \hat V_\text{blue}^{(j)}(t) &= -g_s \sin(\omega t + \phi) \left(e^{-i(4\Omega + \omega) t} \a \mj + e^{i(4\Omega + \omega) t} \c \pj \right) \\
    \hat V_\text{long}^{(j)}(t) &= -g_s \cos(\omega t + \phi) \left(e^{-i(2\Omega + \omega) t} \a + e^{i(2\Omega + \omega) t} \c \right) \Zj,
\end{align}
and those in the charge qubit Hamiltonian are
\begin{align}
    \hat V_{\text{red}, t_c}^{(j)}(t) &= \frac{ig_c}{4}\left((e^{-2i\omega t - i\phi} - 2 e^{-i\omega t}) \a \pj - (e^{2i\omega t + i\phi} - 2 e^{i\omega t}) \c \mj \right) \\
    \hat V_{\text{blue}, t_c}^{(j)} (t) &= i g_c \sin(\frac{\omega t + \phi}{2})^2 \left(e^{-i(4\Omega + \omega)t} \a \mj - e^{i(4\Omega + \omega)t} \c \pj \right) \\
    \hat V_{\text{long}, \epsilon}^{(j)}(t) &= g_c \sqrt{1 - \sin(\frac{\omega t + \phi}{2})^4} \left(e^{-i(2\Omega + \omega)t} \a + e^{i(2\Omega + \omega)t} \c \right) \Zj
\end{align}

\section{Hamiltonian vectorization procedure}

In this section we will describe ideas for efficient bookkeeping in both analytical and numerical computations, which enable us to derive the exact Hamiltonian without RWA in the interaction picture, which is defined by transformation~\eqref{eq: spin unitary transformation} and~\eqref{eq: charge unitary transformation} from the system Hamiltonian.
This will make use of representation theory, namely we will keep track of interaction picture transformations needed for the derivation of the CZ gate which are exactly solvable, i.e. either Gaussian or single-qubit unitaries.

First, we will denote by 
\begin{align}
    \opvec^{(q)} = \frac{1}{\sqrt{2}}\begin{pmatrix} \I^{(q)}, & \X^{(q)}, & \Y^{(q)}, & \Z^{(q)}\end{pmatrix}
\end{align}
the vector collecting the 4 operator basis (where $\I^{(q)}$ is the identity operator in qubit Hilbert space) generators of qubit $q$.
Secondly, we will consider 
\begin{align}
    \opvec^{(0)} = \begin{pmatrix}\id, & \a, & \c, & \c\a \end{pmatrix} 
\end{align}
which will allow us to keep track of transformations on the photon mode.
We will use $\opvec^{(q)}$ and $\opvec^{(0)}$ to express the Hamiltonian of the full system.
To this end, it will be useful to introduce a single vectorization index $\lambda$ which will allow us to keep track of products of operators $\opvec^{(q)}$  on qubit $q$ with operators on the photon.
For any $\mu_0, \mu_1, \mu_2 \in\{0,1,2,3\}$ which label the operator of qubit $q$ and the oscillator, respectively, we define $\lambda(\mu_0, \mu_1, \mu_2)$ in a one-to-one correspondence such that  $\opvec_{\lambda} = \opvec^{(0)}_{\mu_0} \opvec^{(1)}_{\mu_1} \opvec^{(2)}_{\mu_2}$.
For example $\opvec_{\lambda(2,2,0)} = \c \Y^{(1)}$.
In the following, it will be possible to express any operator of interest $\hat O$ as 
\begin{align}
\label{eq:THetaDecomposition}
    \hat O = \sum_\lambda O^\lambda \opvec_{\lambda}. 
\end{align}
In particular, this will be true for the Hamiltonians and for the generators of the interaction picture transformations involved in the derivations.

Next, we will discuss how the relevant transformations can be expressed in the formalism using $\opvec^{(j)}$ and $\opvec^{(0)}$.
Firstly let us consider a rotation $\hat R^{(j)}_\eta(\theta)=\exp\left[-i\theta \hat \sigma^{(j)}_\eta \right]$ for qubit $j$ generated by $\sigma_\mu$ by the angle $\theta$
\begin{align} \label{eq:single_qubit_rotation_Radjoint}
      \hat R^{(j)}_\eta(\theta)^\dagger \sigma^{(j)}_{\eta'}\hat R^{(j)}_\eta(\theta)  =\exp\left[i\theta \hat \sigma^{(j)}_\eta \right] \hat\sigma^{(j)}_{\eta'} \exp\left[-i\theta \hat\sigma^{(j)}_\eta \right] &= \cos(2\theta) \sigma^{(j)}_{\eta'} - \epsilon_{\eta, \eta', \eta''} \sin(2\theta) \hat\sigma^{(j)}_{\eta''} + 2(\sin \theta)^2 \delta_{\eta \eta'} \hat\sigma^{(j)}_\eta 
        \end{align}
When $j=1$, we assign $\eta = \mu$, and we assign $\eta = \mu'$ for $j=2$. Equivalently, we can write this using the $\opvec$ notation as
    \begin{align}
        \hat O =  \langle O,\opvec\rangle
    \end{align}
where $O$ is the vector of coefficients as in Eq.~\eqref{eq:THetaDecomposition} and $\langle\cdot,\cdot\rangle$ implements the summation of the coefficients with the operator entries in the $\opvec^{(j)}$, similarly to the ordinary Euclidean scalar product or Einstein summation convention.
By linearity we have the transformation we have
\begin{align}
    \hat R^{(j)}_\eta(\theta)^\dagger \hat O \hat R^{(j)}_\eta(\theta) 
    =&  \langle \hat O, \opvec^{(o)} \hat R^{(j)}_\eta(\theta)^\dagger \opvec^{(j)} \hat R^{(j)}_\eta(\theta) \opvec^{(j')} \rangle \\
    =&  \langle \hat O, \mathscr{R}^{(j)}_\eta (\theta) \opvec \rangle \\
    =&  \langle \mathscr{R}^{(j)}_\eta(\theta)^T \hat O,\opvec \rangle
    \end{align}
where
\begin{align}
      \mathscr R_0=\id_4,\quad
    \mathscr R_1(\theta) &= 
    \begin{pmatrix}
    1 & 0 & 0 & 0 \\
    0 & 1 & 0 & 0 \\
    0 & 0 & \cos{2\theta} & -\sin{2\theta} \\
    0 & 0 & \sin{2\theta} & \cos{2\theta}
    \end{pmatrix}, \quad
        \mathscr R_2(\theta) &= 
    \begin{pmatrix}
    1 & 0 & 0 & 0 \\
    0 & \cos{2\theta} & 0 & \sin{2\theta} \\
    0 & 0 & 1 & 0 \\
    0 & -\sin{2\theta} & 0 & \cos{2\theta}
    \end{pmatrix},\quad
        \mathscr R_3(\theta) &= 
    \begin{pmatrix}
    1 & 0 & 0 & 0 \\
    0 & \cos{2\theta} & -\sin{2\theta} & 0 \\
    0 & \sin{2\theta} & \cos{2\theta} & 0 \\
     0 & 0 & 0 & 1
    \end{pmatrix}
\end{align}
and 
\begin{align}
    \mathscr R^{(1)}_\mu = \id_o \otimes \mathscr R_\mu \otimes \mathscr R_0, \quad \mathscr R^{(2)}_{\mu'} &= \id_o \otimes \mathscr R_0 \otimes \mathscr R_{\mu'}
\end{align}
This switch from the adjoint action by $\mathscr R^{(j)}_\eta$ on the operators to linear action on the coefficients is equivalent to the exact solution \eqref{eq:single_qubit_rotation_Radjoint}.
It is useful, however, for bookkeping: If we compose rotations then we can simply multiply the corresponding matrices.

Next we consider operations on operators acting on the photon mode using the formalism with $\opvec^{(0)}$.
The transformation $\hat U^{(0)}(\theta) =\exp\left[-i\theta \c\a \right] $
acts as
      \begin{align} \label{eq: photon operator transform}
        \exp\left[i\theta \c\a \right] \a \exp\left[-i\theta \c\a \right] = \a e^{-i\theta}
        =          
        \left(\exp\left[i\theta \c\a \right] \c \exp\left[-i\theta \c\a \right] \right)^\dagger&= \left(\c e^{i\theta}\right)^\dagger
         \end{align}
         and trivially on the occupation number operator
         \begin{align}\exp\left[i\theta \c\a \right] \c\a \exp\left[-i\theta \c\a \right] &= \c\a
\end{align}
This means that given a vector of coefficients $O$ defining the operator $\hat O=\langle O, \opvec^{(0)}\rangle $ acting on the photon mode, we can write its rotation as
\begin{align}
    \hat U^{(0)}(\theta)^\dag \hat O \hat U^{(0)}(\theta) &= \langle \hat O, \hat U^{(0)}(\theta)^\dag\opvec^{(0)}\hat U^{(0)}(\theta) \opvec^{(1)} \opvec^{(2)} \rangle \\
    &= \langle \hat O, \mathscr U^{(0)}(\theta)\opvec \rangle \\
    &= \langle \mathscr U^{(0)}(\theta) \hat O, \opvec \rangle
\end{align}
where
\begin{align}
    \mathscr U^{(0)}(\theta) &= \text{diag}\left(1, e^{-i\theta}, e^{i\theta}, 1 \right) \otimes \id_4 \otimes \id_4 \ .
\end{align}

We then conclude that there are three $4 \times 4$ matrices for the qubit rotation and one $4 \times 4$ for the photon rotation, where the identity generators never changes due to the rotation unitary operators. The photon space generators (identity, number, annihilation, and creation operator) transformation can be described by a diagonal matrix. We are able to conclude the matrices since there are restricted set of generators.

According to the transformation of Hamiltonian:
\begin{align}
    \hat H = \hat U^\dag \left( \hat H_0 - i \frac{\partial}{\partial t} \right) \hat U
\end{align}

we are able to write the superoperator form of the time-dependent unitary transformation:
\begin{align}
    \mathscr{U}(t; \omega_r, \omega_j, \omega_k) \hat H = \mathscr U^{(0)} (\omega_r t) \mathscr R_{\mu}^{(j)} (\omega_j t) \mathscr R_{\mu'}^{(k)} (\omega_k t) \hat H - \omega_r \opvec^{(0)}_{3} - \omega_j \opvec^{(j)}_{\mu} - \omega_k \opvec^{(k)}_{\mu'},
\end{align}
whereas the time-independent transformations take a simpler form:
\begin{align}
    \mathscr{U}(\theta_0, \theta_j, \theta_k) \hat H = \mathscr U^{(0)} (\theta_0) \mathscr R_{\mu}^{(j)} (\theta_j) \mathscr R_{\mu'}^{(k)} (\theta_k) \hat H
\end{align}
It can be shown that we can focus on the coupled-qubit subspace of the $\hat H_0$ since after the first transformation $\hat U_1(t) = e^{-i \omega t \left( \c\a + \sum_q \frac{1}{2}\Z^{(q)}\right)}$, the idle qubit Hamiltonian is rotated away.

We can therefore write the transformation superoperator acting on the vectorized elements of $\hat H_0^{(j)}$ in
\begin{align}
    \hat H = \mathscr{U}\left(0, -\frac{\pi}{4}, -\frac{\pi}{4}\right) \mathscr{U}\left(t; 2\Omega, \Omega, 0\right) \mathscr{U}\left(0, \frac{\phi}{2} + \frac{\pi}{4}, \frac{\phi}{2} + \frac{\pi}{4}\right) \mathscr{U}\left(t; \omega, \omega, \omega\right) \hat H_0^{(j)}
\end{align}

After all four unitary transformations, we want to represent the vectorized operator in the new qubit generators defined by:
\begin{align*}
    \opvec^{(q)}_{+} + \opvec^{(q)}_{-} &= \opvec^{(q)}_{1} \\
    -i \opvec^{(q)}_{+} + i\opvec^{(q)}_{-} &= \opvec^{(q)}_{2}
\end{align*}
in which $\opvec^{(q)}_{\{+, -\}} = \frac{1}{\sqrt{2}} \sigma^{(q)}_{\{+, -\}}$, and the zeroth and third generators are unchanged. It can be easily deduced that the qubit-subspace operator under the new generators representation can be calculated by:
\begin{align}
    \begin{pmatrix} O'^{0}\\ O'^{1} \\ O'^{2} \\ O'^{3} \end{pmatrix} = \begin{pmatrix} 1&0&0&0\\ 0 & 1 & -i & 0 \\ 0 & 1 & i & 0 \\ 0&0&0&1 \end{pmatrix} \begin{pmatrix} O^{0}\\ O^{1} \\ O^{2} \\ O^{3} \end{pmatrix}
\end{align}
where elements $O'^\eta$ in the $\opvec^{(q)}_{\{0, +, -, 3\}}$ generator can be expressed by $O^\eta$, elements of $\opvec^{(q)}_{\{0, 1, 2, 3\}}$ generator.
\begin{align} \label{eq: vector reprezetaentation of exact Hamiltonian}
    \hat H_R(\zeta) &= \sum_\lambda h_R^\lambda(\zeta) \, \opvec_\lambda \\
    \hat H_E(\zeta;t) &= \sum_\lambda h_E^\lambda(\zeta;t) \, \opvec_\lambda
\end{align}

\section{\label{sec: fourier decomposition} Fourier decomposition of exact Hamiltonians}
The exact Hamiltonian~\eqref{eq: spin exact Hamiltonian} and~\eqref{eq: charge exact Hamiltonian} can be decomposed by complex periodic function $e^{i \varpi t}$ at angular frequency $\varpi$. Using the vectorized formulation~\eqref{eq: vector reprezetaentation of exact Hamiltonian}, we are able to write $h^\lambda_E(\zeta;\tau) = h^\lambda(\zeta; 0) + \sum_{\tilde\varpi} h^\lambda_E(\zeta; i \tilde\varpi \tau)$, where $\tilde\varpi$ denote all nonzero frequencies. $h^\lambda(\zeta; 0)$ denotes the constant terms in the Hamiltonian. 

In the following discussion, the vectorization index $\lambda(\mu_0, \mu_1, \mu_2)$ are adjusted to $\lambda(\mu_0, \mu_j)$ where $j$ labels the coupled qubit, since the uncoupled qubit does not contribute terms in the interaction picture.

In our case, they are $h_E^{\lambda(1,+)}(\zeta; 0) = \frac{ie^{i\phi}}{2}$ and $h_E^{\lambda(2,-)}(\zeta; 0) = -\frac{ie^{-i\phi}}{2}$ for both spin and charge qubit exact Hamiltonians.

For the spin qubit Hamiltonian, all Fourier components can be found by expanding the sine and cosine functions, e.g.,
\begin{align}
    h_E^{\lambda(1,-)}(\zeta; \tau) &= -\sin(\tilde\omega\tau + \phi) e^{-i(4\tilde\Omega + \tilde\omega) \tau} = \frac{ie^{i\phi}}{2} e^{-i4\tilde\Omega\tau} - \frac{ie^{-i\phi}}{2} e^{-i(4\tilde\Omega + 2\tilde\omega)\tau} \nonumber \\
    &= h_E^{\lambda(1,-)}(\zeta; -i4\tilde\Omega\tau) + h_E^{\lambda(1,-)}(\zeta; -i(4\tilde\Omega + 2\tilde\omega)\tau) \nonumber
\end{align}

The terms in $\hat V_{\text{qubit}}(t) = \sum_{\mu_j \neq 0} h_E^{\lambda(0,\mu_j)}(\zeta; \tau) \opvec_{\lambda(0,\mu_j)}$ are proportional to $\frac{\tilde\Omega}{2}$ for $\tilde\varpi \in \pm \{2\tilde\omega, 2(\tilde\omega + \tilde\Omega), 2(\tilde\omega - \tilde\Omega)\}$. The rest of the terms have non-zero frequencies $\tilde\varpi \in \pm \{2\tilde\Omega, 4\tilde\Omega, 2(\tilde\omega + \tilde\Omega), 2(\tilde\omega + 2\tilde\Omega), 2\tilde\omega\}$.

The Fourier components of the charge qubit Hamiltonian can be found in the same way except $h_E^{\lambda(1,3)}$ and $h_E^{\lambda(2,3)}$ corresponding to the longitudinal interaction $\hat V_{\text{long}, \epsilon}(t)$. The $\epsilon(t)$ drive can be decomposed by
\begin{align} \label{eq: epsilon fourier expansion}
    \sqrt{1 - \sin(\frac{\omega t + \phi}{2})^4} &= \sum_{u=0}^\infty \frac{(-1)^u \sqrt{\pi}}{2 u! \Gamma\left(\frac{3}{2} - u\right)} \left(\frac{1}{2i}\right)^{4u} \left(e^{\frac{i\omega t + i\phi}{2}} - e^{-\frac{i\omega t + i\phi}{2}}\right)^{4u} \nonumber \\ 
    &= \sum_{u=0}^\infty \sum_{v=0}^{4u} \frac{(-1)^{u+v} \sqrt{\pi} C^{4u}_v}{2 u! \Gamma\left(\frac{3}{2} - u\right) (2i)^{4u}} \, e^{i(2u-v)(\omega t + \phi)} \\
    &= \sum_{u=0}^\infty \sum_{v=0}^{4u} E_{uv} \, e^{i(2u-v)(\omega t + \phi)} \nonumber
\end{align}
wherein the relationship between the generalized binomial coefficient and the gamma function $C^{\frac{1}{2}}_n = \frac{\Gamma\left(\frac{3}{2}\right)}{\Gamma\left(n+1\right)\Gamma\left(\frac{3}{2} - n\right)}$ has been used.

The longitudinal terms contain constant when $\tilde\Omega = \frac{\tilde\omega_r}{2}\left(1 - \frac{1}{p}\right)$, where $p = 2u - v$ and $-2u \leq p \leq 2u, p \neq 0$. Therefore, we denote the longitudinal interaction Hamiltonian $\hat V_{\text{long}, \epsilon} = (D(\tau)\a + D(\tau)^* \c)\Zj$, where $D(\tau) = 2\sqrt{1 - \sin(\frac{\tilde\omega \tau + \phi}{2})^4} e^{-i\tilde\omega_r \tau}$. The longitudinal interaction with frequency $\varpi$ is then
\begin{align}
    h_E^{\lambda(1,3)}(\zeta; i\tilde\varpi\tau) &= \left(2e^{ip\phi}\sum_{p\tilde\omega-\tilde\omega_r = \tilde\varpi} E_{uv} \right) e^{i\tilde\varpi\tau} \nonumber \\
    h_E^{\lambda(2,3)}(\zeta; i\tilde\varpi\tau) &= \left(2e^{ip\phi}\sum_{p\tilde\omega+\tilde\omega_r = \tilde\varpi} E_{uv} \right) e^{i\tilde\varpi\tau} \nonumber
\end{align}
It is noted that the magnitude of the large $\tilde\varpi$ ($p$ is large) is negligible compared to that of the small $\tilde\varpi$ ($p$ is small), since $C^{4u}_{2u-p_1} \ll C^{4u}_{2u-p_2}$ if $p_1 > p_2$ when $u$ is large. The terms with $u_1$ is much smaller than that with $u_2$ when $u_1 > u_2$ since $2^{-4u_1} \ll 2^{-4u_2}$ when $u_1, u_2$ is large.

\section{Analytical approximation for fidelity and high-fidelity conditions}
The fidelity of the exact unitary evolution can be approximated by truncating the photon Fock space at $N$,
\begin{align} \label{eq: unitary fidelity}
    F = \frac{1}{4N} \Bigg|\Tr\Bigg[ \Bigg( \mathcal{T}_{\rightarrow} \prod_\zeta \exp\Bigg[ i \int_{\tiz}^{\tfz} d\tau \HR(\zeta; \tau) \Bigg] \hat Z^\dag \hat M_0 \Bigg) \Bigg( \hat M_0 \hat Z \mathcal{T}_{\leftarrow} \prod_\zeta  \exp\Bigg[ -i \int_{\tiz}^{\tfz} d\tau \HE(\zeta; \tau) \Bigg] \Bigg) \Bigg] \Bigg|
\end{align}

Calculating $\hat Z^\dag \hat M_0 \hat Z = \ket{0}\bra{0} \otimes \hat \sigma_0 \otimes \hat \sigma_0 $, we can write
\begin{align}
    F = \frac{1}{4} \Bigg|\Tr\bra{0} \SE(5) \SE(4) \SE(3) \SE(2) \SE(1) \SR(1)^\dagger \SR(2)^\dagger \SR(3)^\dagger \SR(4)^\dagger \SR(5)^\dagger \ket{0}\Bigg|.
\end{align}
Expanding each sideband transition according to Neumann series until the second order,
\begin{align}
    \SR(\zeta; \Delta\tau_\zeta, \phi_\zeta)^\dagger &\approx \id + i\int_{\tiz}^{\tfz} d\tau \HR(\zeta) -\frac{1}{2} \left(\int_{\tiz}^{\tfz} d\tau \HR(\zeta)\right)^2 \nonumber \\
    &= \id + i\hat I_{R1}(\zeta) - \hat I_{R2}(\zeta) \\
    \SE(\zeta; \Delta\tau_\zeta, \phi_\zeta) &\approx \id - i\int_{\tiz}^{\tfz} d\tau \HE(\zeta; \tau) - \int_{\tiz}^{\tfz} d\tau_1 \int_{\tiz}^{\tau_1} d\tau_2 \, \HE(\zeta; \tau_1) \HE(\zeta; \tau_2) \nonumber \\
    &= \id - i \hat I_{E1}(\zeta) - \hat I_{E2}(\zeta)
\end{align}
the analytical approximation to fidelity up to the second order is
\begin{align} \label{eq: second order fidelity}
    F_2 = \Bigg| 1 + \frac{1}{4}\Tr\bra{0}\Bigg( - \sum_\zeta \hat I_{R2}(\zeta) - \sum_\zeta \hat I_{E2}(\zeta) - \sum_{\zeta < \zeta'} \hat I_{R1}(\zeta) \hat I_{R1}(\zeta') - \sum_{\zeta > \zeta'} \hat I_{E1}(\zeta) \hat I_{E1}(\zeta') + \sum_{\zeta \zeta'} \hat I_{E1}(\zeta) \hat I_{R1}(\zeta')\Bigg)\ket{0} \Bigg|.
\end{align}
We gain insight from calculating the first order and second exact Hamiltonian integrals,
\begin{align}
    \hat I_{E1}(\zeta) &= \sum_\lambda \opvec_\lambda \int_{\tiz}^{\tfz} d\tau \, h^\lambda_E(\zeta;\tau) = \sum_\lambda \opvec_\lambda I^\lambda_{E1}(\zeta), \\
    \hat I_{E2}(\zeta) &= \sum_{\lambda \lambda'} \opvec_\lambda \opvec_{\lambda'} \int_{\tiz}^{\tfz} d\tau_1 \int_{\tiz}^{\tau_1} d\tau_2 \, h^{\lambda}_E(\zeta;\tau_1) h^{\lambda'}_E(\zeta;\tau_2) = \sum_{\lambda \lambda'} \opvec_\lambda \opvec_{\lambda'} I^{\lambda \lambda'}_{E2}(\zeta).
\end{align}
Note that for the Hamiltonians, $h^\lambda(\zeta; 0) = h^\lambda_R(\zeta)$ when $\lambda = 5,10$, and $h^\lambda(\zeta; 0) = 0$ otherwise. Therefore, the integrals have the form
\begin{align}
    I^\lambda_{R1}(\zeta) = &h^\lambda(\zeta; 0) \Delta\tau_\zeta \\
    I^{\lambda \lambda'}_{R2}(\zeta) = &\frac{\Delta\tau_\zeta^2}{2} h^\lambda(\zeta; 0) h^{\lambda'}(\zeta; 0) = \frac{1}{2} I^\lambda_{R1}(\zeta) I^{\lambda'}_{R1}(\zeta)\\
    I^\lambda_{E1}(\zeta) = &h^\lambda(\zeta; 0) \Delta\tau_\zeta + \sum_{\tilde\varpi} \frac{h^\lambda_E(\zeta; i \tilde\varpi \tau)}{i \tilde\varpi} \Big|_{\tiz}^{\tfz} \\
    I^{\lambda \lambda'}_{E2}(\zeta) = &\frac{\Delta\tau_\zeta^2}{2} h^\lambda(\zeta; 0) h^{\lambda'}(\zeta; 0) + h^\lambda(\zeta; 0) \sum_{\tilde\varpi'} \left(\frac{h^{\lambda'}_E(\zeta; i\tilde\varpi'\tau)}{(i\tilde\varpi')^2}\Big|_{\tiz}^{\tfz} - \frac{h^{\lambda'}_E(\zeta; i\tilde\varpi'\tiz) \Delta\tau_\zeta}{i\tilde\varpi'}\right) \nonumber \\
    &+ \sum_{\tilde\varpi} \left(\frac{h^\lambda_E(\zeta; i\tilde\varpi\tau)}{\tilde\varpi^2}\Big|_{\tiz}^{\tfz} + \frac{h^\lambda_E(\zeta; i\tilde\varpi\tfz) \Delta\tau_\zeta}{i\tilde\varpi}\right) h^{\lambda'}(\zeta; 0) \nonumber \\
    &+ \sum_{\tilde\varpi \tilde\varpi'} \frac{1}{i\tilde\varpi'} \left(\frac{h^{\lambda}_E(\zeta; i\tilde\varpi\tau) h^{\lambda'}_E(\zeta; i\tilde\varpi'\tau)}{i(\tilde\varpi + \tilde\varpi')}\Big|_{\tiz}^{\tfz} - \frac{h^{\lambda}_E(\zeta; i\tilde\varpi\tau)h^{\lambda'}_E(\zeta; i\tilde\varpi'\tiz)}{i\tilde\varpi}\Big|_{\tiz}^{\tfz}\right)
\end{align}

For high fidelity, we observe from the integrals that all the frequency $\tilde\varpi$-dependent terms should be vanishingly small, so the high fidelity condition from the first order integral is $\abs{\tilde\varpi} \gg \abs{h^\lambda_E(\zeta; i \tilde\varpi \tau)}$. From the second order integral, the disappearance of the $\tilde\varpi, \tilde\varpi'$-dependent terms demonstrate more implicit conditions for high fidelity, which means the fidelity varies swiftly when we change the frequency parameters $\Omega$ and $\omega_r$. However, a straightforward way to make them disappear is $\abs{\tilde\varpi}, \abs{\tilde\varpi'} \gg \abs{h^\lambda_E(\zeta; i \tilde\varpi \tau)}, \abs{h^{\lambda'}_E(\zeta; i \tilde\varpi' \tau)}, \Delta\tau_s$.

In the exact Hamiltonians for both systems, the smallest non-zero frequency is $2\tilde \Omega$ in the resonant regime. To make the term $\abs{\frac{h^\lambda_E(\zeta; i2\tilde\Omega\tfz) \Delta\tau_s}{i2\tilde\Omega}}$ smaller than some small threshold, say $1/8$, we demand $\tilde\Omega > \max\Delta\tau_s = \sqrt{2}\pi \approx 4.44$. This defines a lower boundary of the high fidelity region across all $\tilde\Omega$.

The RWA Hamiltonian shares the same first order integrals
\begin{align*}
    I^{0}_{R}(\zeta) &= \frac{i \tau e^{i \phi}}{2}\Bigg|^{\tfz}_{\tiz} \\
    I^{1}_{R}(\zeta) &= - \frac{i \tau e^{- i \phi}}{2}\Bigg|^{\tfz}_{\tiz}
\end{align*}

For the spin qubit Hamiltonian, the first order integrals are
\begin{align}
    I^{1}_{E1}(\zeta) &= \frac{\tilde\Omega \left(- i \tilde\Omega \sin{\left(2 \tilde\omega \tau + 2 \phi \right)} + 2 \omega \cos{\left(2 \tilde\omega \tau + 2 \phi \right)}\right) e^{i \tilde\Omega \tau}}{\tilde\Omega^{2} - 4 \tilde\omega^{2}}\Bigg|^{\tfz}_{\tiz}  \label{eq: 1st-order integrals 1}\\
    I^{2}_{E1}(\zeta) &= \frac{\tilde\Omega \left(i \tilde\Omega \sin{\left(2 \tilde\omega \tau + 2 \phi \right)} + 2 \tilde\omega \cos{\left(2 \tilde\omega \tau + 2 \phi \right)}\right) e^{- i \tilde\Omega \tau}}{\tilde\Omega^{2} - 4 \tilde\omega^{2}}\Bigg|^{\tfz}_{\tiz} \label{eq: 1st-order integrals 2}\\
    I^{3}_{E1}(\zeta) &= \frac{\tilde\Omega \sin{\left(2 \tilde\omega \tau + 2 \phi \right)}}{2 \tilde\omega}\Bigg|^{\tfz}_{\tiz} \label{eq: 1st-order integrals 3}\\
    I^{5}_{E1}(\zeta) &= \frac{i \tau e^{i \phi}}{2} + \frac{e^{- 2 i \tilde\omega \tau - i \phi}}{4 \tilde\omega}\Bigg|^{\tfz}_{\tiz} \label{eq: 1st-order integrals 5}\\
    I^{6}_{E1}(\zeta) &= - \frac{\left(4 i \tilde\Omega \sin{\left(\tilde\omega \tau + \phi \right)} + i \tilde\omega \sin{\left(\tilde\omega \tau + \phi \right)} + \tilde\omega \cos{\left(\tilde\omega \tau + \phi \right)}\right) e^{- i \tau \left(4 \tilde\Omega + \tilde\omega\right)}}{8 \tilde\Omega \left(2 \tilde\Omega + \tilde\omega\right)}\Bigg|^{\tfz}_{\tiz} \label{eq: 1st-order integrals 6}\\
    I^{7}_{E1}(\zeta) &= \frac{\left(- 2 i \tilde\Omega \cos{\left(\tilde\omega \tau + \phi \right)} + \tilde\omega \sin{\left(\tilde\omega \tau + \phi \right)} - i \tilde\omega \cos{\left(\tilde\omega \tau + \phi \right)}\right) e^{- i \tau \left(2 \tilde\Omega + \tilde\omega\right)}}{4 \tilde\Omega \left(\tilde\Omega + \tilde\omega\right)}\Bigg|^{\tfz}_{\tiz} \label{eq: 1st-order integrals 7}\\
    I^{9}_{E1}(\zeta) &= \frac{\left(4 i \tilde\Omega \sin{\left(\tilde\omega \tau + \phi \right)} + i \tilde\omega \sin{\left(\tilde\omega \tau + \phi \right)} - \tilde\omega \cos{\left(\tilde\omega \tau + \phi \right)}\right) e^{i \tau \left(4 \tilde\Omega + \tilde\omega\right)}}{8 \tilde\Omega \left(2 \tilde\Omega + \tilde\omega\right)}\Bigg|^{\tfz}_{\tiz} \label{eq: 1st-order integrals 9}\\
    I^{10}_{E1}(\zeta) &= - \frac{i\tau e^{- i \phi}}{2} + \frac{ e^{2 i \tilde\omega \tau + i\phi}}{4 \tilde\omega}\Bigg|^{\tfz}_{\tiz} \label{eq: 1st-order integrals 10}\\
    I^{11}_{E1}(\zeta) &= \frac{\left(2 i \tilde\Omega \cos{\left(\tilde\omega \tau + \phi \right)} + \tilde\omega \sin{\left(\tilde\omega \tau + \phi \right)} + i \tilde\omega \cos{\left(\tilde\omega \tau + \phi \right)}\right) e^{i \tau \left(2 \tilde\Omega + \tilde\omega\right)}}{4 \tilde\Omega \left(\tilde\Omega + \tilde\omega\right)}\Bigg|^{\tfz}_{\tiz} \label{eq: 1st-order integrals 11}
\end{align}
with $\omega = \omega_r - 2\Omega$, and the evaluation of upper and lower limit act on the expression to its left side.

To roughly estimate the upper boundary of the high-fidelity region, we first observe the first two exact integrals and try to set the their product smaller than a threshold, which we take as $1/2^9$:
\begin{align} \label{eq: ratio}
    \frac{\tilde\Omega^4 + 4\tilde\omega^2 \tilde\Omega^2}{(\tilde\Omega^2 - 4\tilde\omega^2)^2} < \frac{1}{2^9},
\end{align}
since the other ratios are definitely smaller than this one for positive $\tilde \omega_r$ and $\tilde \Omega$. We solve for $\tilde \omega_r/\tilde \Omega$, and obtain $\tilde \omega_r/\tilde \Omega > \frac{1}{2}(4 + \sqrt{257+ 32\sqrt{65}}) \approx 13.347$. By observing the numerical evidence, we conclude the high-fidelity condition: a lower and upper boundary define a high-fidelity plateau.
\begin{align} \label{eq: spin-qubit high-F condition}
    \tilde\Omega \in \left(\sqrt{2}\pi, \frac{\tilde\omega_r}{13.347}\right)
\end{align}

For the charge qubit, the longitudinal terms contain a constant at $\tilde\Omega = \frac{\tilde\omega_r}{2}\left(1 - \frac{1}{p}\right)$ as discussed before. The unwanted constant terms diminishes the fidelity at these values. Additionally, when $\tilde\Omega = \frac{\tilde\omega_r}{2}$, the driving frequency $\tilde\omega = 0$, so that the red sideband transition terms vanish in stage $\zeta = 1,2,4,5$. The evaluation of the second-order integral gives the same lower limit of $\tilde\Omega$ as it is for the spin qubit. Therefore, the high-fidelity condition of the charge qubit is
\begin{align}
    \tilde\Omega \in \mathbbm{R}^{+} \setminus \Biggl\{ [0, \frac{\sqrt{2}\pi}{2}), \frac{\tilde\omega_r}{2}, \frac{\tilde\omega_r}{2}\left(1 - \frac{1}{p}\right) \Biggr\}
\end{align}

\section{Quantum gate with decoherence: superoperator calculations}
In this section we will discuss superoperators relevant for open system dynamics.
Again, we will use similar bookkeeping techniques as previously discussed.

In the protocol for the CZ gate, there are 7 stages ordered by: five of them denoted by $\zeta$ is the sideband transition evolution; one of two-qubit rotation and the final projection to 0-photon subspace.
For example, the unitary evolution $\hat U_\text{CZ}$ of the density matrix $\hat\rho = \hat U_\text{CZ} \hat\rho(0) \hat U_\text{CZ}^{\dag}$ of the ideal CZ $=\hat U_{cz}$ gate will be denoted as
\begin{align} \label{eq: ideal gate superoperator}
    \hat \rho &= \mathscr{U}_\text{CZ} \hat \rho(0) \nonumber \\
    &= \mathscr M_0 \mathscr Z \mathcal{T}_{\leftarrow} \prod_{\zeta = 1}^{5} \mathscr U(\zeta; \tfz, \tiz) \nonumber\\
    &= \mathscr M_0 \mathscr Z \mathscr U(5; \tau_{\text{f}5}, \tau_{\text{i}5}) \mathscr U(4; \tau_{\text{f}4}, \tau_{\text{i}4}) \mathscr U(3; \tau_{\text{f}3}, \tau_{\text{i}3}) \mathscr U(2; \tau_{\text{f}2}, \tau_{\text{i}2}) \mathscr U(1; \tau_{\text{f}1}, \tau_{\text{i}1})
\end{align}
wherein
\begin{align}
    \mathscr M_0 \hat\rho &= \hat M_0 \rho \hat M_0 \label{eq: projection superoperator}\\
    \mathscr Z \hat\rho &= \hat Z^{(1)} \left( \frac{\pi}{\sqrt{2}} \right) \hat Z^{(2)} \left(- \frac{\pi}{\sqrt{2}}\right) \hat\rho \hat Z^{(2)} \left(- \frac{\pi}{\sqrt{2}}\right)^\dag \hat Z^{(1)} \left( \frac{\pi}{\sqrt{2}} \right)^\dag \label{eq: Z superoperator}\\
    \mathscr U(\zeta; \tfz, \tiz) \hat\rho &= \SR(\zeta; \Delta\tau, \phi) \hat\rho \, \SR(\zeta; \Delta\tau, \phi)^\dag
\end{align}

We begin the discussion of open system dynamics in Schr\"odinger picture where in general the Lindblad master equation reads~\cite{breuer2002theory}
\begin{align}
    \frac{\mathrm d}{\mathrm dt} \hat\rho(t) = \mathscr{L} \hat\rho(t) = -i \comm{\hat H}{\hat\rho(t)} + \sum_{\ell} \gamma_\ell \left( \hat L_\ell \hat\rho(t) \hat L^{\dag}_\ell - \frac{1}{2} \acomm{\hat L^{\dag}_\ell  \hat L_\ell}{\hat \rho(t)} \right)
\end{align}
For the system at hand we model qubit dephasing using the Lindblad operator $\hat L_{1,2} = \Z^{(1,2)}$ and take rates  $\gamma_1 = \gamma_2 = \gamma$  to be the same for both qubits. In turn, for the photon mode, we will take the rate $\gamma_3 =\kappa$ and Lindblad operator $\hat L_3 = \a$.
Thus, explicitly the Lindblad master equation for the system will be
\begin{align}
    \frac{\mathrm d}{\mathrm dt} \hat\rho(t) = -i \comm { \hat H_E}{\hat\rho(t)} + \sum_{q=1,2} \gamma \left(\Z^{(q)} \hat\rho \Z^{(q)} - \hat\rho \right) + \kappa \left(\a \hat\rho \c - \frac{1}{2} \c\a \hat\rho - \frac{1}{2} \hat\rho \c\a\right).
\end{align}
Transforming the master equation into the corresponding interaction picture with stage index {$\zeta$}:
\begin{align} \label{eq: Lindblad equation}  
    \frac{\mathrm d}{\mathrm d \tau} \hat\rho(\tau) = \mathscr{L} (\zeta; \tau) \hat\rho(\tau) = -i \comm{\HE(\zeta; \tau)}{\hat\rho(\tau)} + \sum_{\ell} \tilde \gamma_\ell \left( \hat L_\ell(\zeta; \tau) \hat\rho(\tau) \hat L_\ell(\zeta; \tau)^{\dag} - \frac{1}{2} \acomm{\hat L_\ell(\zeta; \tau)^{\dag}  \hat L_\ell(\zeta; \tau)}{\hat \rho(\tau)} \right)
\end{align}
where for spin qubit system, using indices $j, k$ to indicate coupled and uncoupled qubits,
\begin{align} \label{eq: Lindblad operator for spin qubit}
    \hat L_{j}(\zeta; \tau) &=  - \sin(2 \tilde \Omega \tau) \X^{(j_\zeta)} - \cos(2\tilde \Omega \tau) \Y^{(j_\zeta)} \\
    \hat L_{k}(\zeta; \tau) &= -\Y^{(k_\zeta)} \\
    \hat L_3 (\tau) &= e^{-i(\tilde \omega + 2\tilde \Omega)\tau} \a
\end{align}
and charge qubit system:
\begin{align} \label{eq: Lindblad operator for charge qubit}
    \hat L_{j}(\zeta; \tau) &= \Z^{(j)} \\
    \hat L_{k}(\zeta; \tau) &= \Z^{(k)} \\
    \hat L_3 (\tau) &= e^{-i(\tilde \omega + \tilde \Omega)\tau} \a
\end{align}

The dissipation rates in the Lindblad master equation are rescaled against the coupling constant: $ \gamma/g \rightarrow \tilde \gamma $, and $ \kappa/g \rightarrow \tilde \kappa $.
Next, we formulate the notation for a general solution~\cite{breuer2002theory} of a time-dependent Lindblad master equation~\eqref{eq: Lindblad equation} in stage $\zeta$: 
\begin{align} \label{eq: exact gate superoperator}
    \hat \rho (\tfz) = \mathscr{S}(\zeta; \tfz, \tiz) \hat \rho (\tiz) &= \mathcal{T}_{\leftarrow} \exp \left[ \int_{\tiz} ^{\tfz} d\tau \mathscr{L}(\zeta; \tau) \right] \hat \rho (\tiz) \nonumber \\
    &\approx \mathcal{T}_{\leftarrow} \prod_{\xi=0}^{T-1} \exp\left[ d\tau \mathscr{L}(\zeta; \tau_\xi)\right]
\end{align}
wherein $T-1$ is the total number of discretized time steps, $\xi$ is the time increment index, and $d\tau = \frac{\tfz- \tiz}{T-1} $ is the increment magnitude. The approximate equal becomes exact when $T \rightarrow \infty$. $\mathscr{S}(s; \tf, \ti)$, which is independent of initial states, and defines a smooth semigroup evolution from an arbitrary density matrix for stage $\zeta$, from which we can find the Markovian, dissipative quantum gate.

Now we connect the CZ gate protocol to the open system discussion by replacing the unitary evolution superoperator $\mathscr U(s; \tfz, \tiz)$ by semigroup evolution $\mathscr S(\zeta; \tfz, \tiz)$:
\begin{align} \label{eq: semigroup evolution CZ gate}
    \mathscr{S}_{cz} = \mathscr{M}_0 \mathscr{Z} \left( \mathcal T_{\leftarrow}\prod_{\zeta=0}^{4} \mathscr{S}(\zeta; \tfz, \tiz) \right).
\end{align}

The vectorized superoperators can be written in matrix form~\cite{greenbaum2015introduction}, and the elements are found under set of generators $\opvec_{\lambda} = \opvec^{(o)}_{\mu_o} \opvec^{(1)}_{\mu_1} \opvec^{(2)}_{\mu_2}$, where $\opvec^{(o)}_{\mu_o} = \ket{m}\bra{n} \otimes \hat \sigma_0 \otimes \hat \sigma_0 $, and $\opvec^{(q=1,2)} = \frac{1}{\sqrt{2}}\begin{pmatrix} \I^{(q)}, & \X^{(q)}, & \Y^{(q)}, & \Z^{(q)} \end{pmatrix}$.

In general, a superoperator $\mathscr O$ acting on the density matrix can be represented by a matrix with elements:
\begin{align}
    \tensor{\mathscr{O}}{^\mu_\nu}  = \text{Tr} \left[ \opvec_\mu^\dagger \mathscr{O} \opvec_\nu \right].
\end{align}

For a continuous time-dependent (described by time $t$) and composite superoperator that can be decomposed into summation of terms that consist a time-dependent function and time-independent joint superoperators (but may depend on stages $\zeta$) act on each subspace labeled by $m \in \{0, 1, 2\}$ for oscillator and two qubits
\begin{align}
    \mathscr{O}(\zeta; t) &= \sum_{u} f_u (\zeta; t) \mathscr{O}_u(\zeta) = \sum_{u = 0} f_u (\zeta; t) \prod_{m} \mathscr{O}^{(m)}_u(\zeta) \\
    &= \sum_{u} f_u (\zeta; t)  \mathscr{O}_u^{(0)}(\zeta)\mathscr{O}_u^{(1)}(\zeta)\mathscr{O}_u^{(2)}(\zeta) .
\end{align}

The decomposed superoperator $\mathscr{O}_u(\zeta)$ has the tensor product factor $\mathscr{O}^{(\zeta)}_u(\zeta)$, which satisfy $\comm{\mathscr{O}^{(m)}_u(\zeta)}{\mathscr{O}^{(m')}_{u'}(\zeta')}\Big|_{m \neq m'} = 0$ since they act on different subspace. Therefore, we can write the superoperator tensor product factors $\mathscr{O}^{(m = 0)}_u(\zeta)$, $\mathscr{O}^{(m = 1)}_u(\zeta)$, and $\mathscr{O}^{(m = 2)}_u(\zeta)$ in any order without lossing generality.

In the photon-qubits system, we have the generator set
\begin{align}
    \opvec_\lambda = \opvec^{(0)}_{\mu_0} \opvec^{(1)}_{\mu_1} \opvec^{(2)}_{\mu_2}
\end{align}
where the index in each subspace $\mu_m \in \{0, 1, \ldots, d_m^2\}$.

The elements of a superoperator can then be found:
\begin{align} \label{eq: superoperator elements}
    \tensor{\mathscr{O}(\zeta; t)}{^{\mu_o \mu_1 \mu_2}_{\nu_o \nu_1 \nu_2}} &= \sum_u \Tr \left[ f_u (\zeta; t) \, \opvec^{(0) \dagger}_{\mu_o} \mathscr{O}_{u}^{(0)}(\zeta) \opvec^{(0)}_{\nu_o} \, \opvec^{(1)\dagger}_{\mu_1} \mathscr{O}_{u}^{(1)}(\zeta) \opvec^{(1)}_{\nu_1} \, \opvec^{(2) \dagger}_{\mu_2} \mathscr{O}_{u}^{(2)}(\zeta) \opvec^{(2)}_{\nu_2} \right] \nonumber \\
    \tensor{\mathscr{O}(\zeta; t)}{^\lambda_{\lambda'}} &= \sum_u f_u (\zeta; t) \, \tensor{\mathscr{O}_{u}^{(0)}(\zeta)}{^{\mu_0}_{\nu_0}} \, \tensor{\mathscr{O}_{u}^{(1)}(\zeta)}{^{\mu_1}_{\nu_1}} \, \tensor{\mathscr{O}_{u}^{(2)}(\zeta)}{^{\mu_2}_{\nu_2}}.
\end{align}

By using Eq.~\eqref{eq: superoperator elements}, we can find the matrix representation of Lindblad superoperator $\mathscr L(\zeta; t)$ and therefore obtain the semigroup evolution superoperator by Eq.~\eqref{eq: semigroup evolution CZ gate}.

Finally we check the fidelity~\cite{wang2008alternative} of the protocol using: 
\begin{align}
    \mathcal F(\tilde\Omega, \tilde\omega_r, \tilde\gamma, \tilde\kappa) &= \frac{1}{d^2} \abs{\Tr \left[\mathscr{U}_{cz}^{\dag} \mathscr{S}_{cz} \right]}
    \label{eq: superoperator fidelity} \\
    F(\tilde\Omega, \tilde\omega_r) &= \frac{1}{d} \abs{\Tr \left[U_{cz}^{\dag} U_{E} \right]}
    \label{eq: fidelity}
\end{align}
where $d=4$ is the dimension of the two-qubit Hilbert space. When $\mathscr{S}_{cz} = \mathscr{U}_{cz}$, i.e., both of them are unitary, the fidelity is $1$. When the $\mathscr{S}_{cz}$ is completely orthogonal to $\mathscr{U}_{cz}$, the fidelity is $0$, which meets the fundamental requirements to define a measure of quantum operator fidelity. The same is true for operator fidelity $F$.

\section{Numerical method}
In the numerical evaluation of semigroup evolution in Eq.~\eqref{eq: semigroup evolution CZ gate}, to ensure convergence, the total number of time steps assigned is based on the maximum frequency of oscillation in the evolution. We assign 64 time instances for the shortest period of oscillation so that for each stage $\zeta$, the total number of discretized time instances is:
\begin{align} \label{eq: numerical samples number}
    T = \left\lceil \frac{\max(\Delta \tau_\zeta) \tilde \omega_{\text{max}}}{2 \pi} \times 64 + 1 \right\rceil,
\end{align}
where the ceiling notation $\lceil ~\rceil$ rounds up to the nearest integer. 

\begin{figure}
    \centering
    \includegraphics[width=0.4\linewidth]{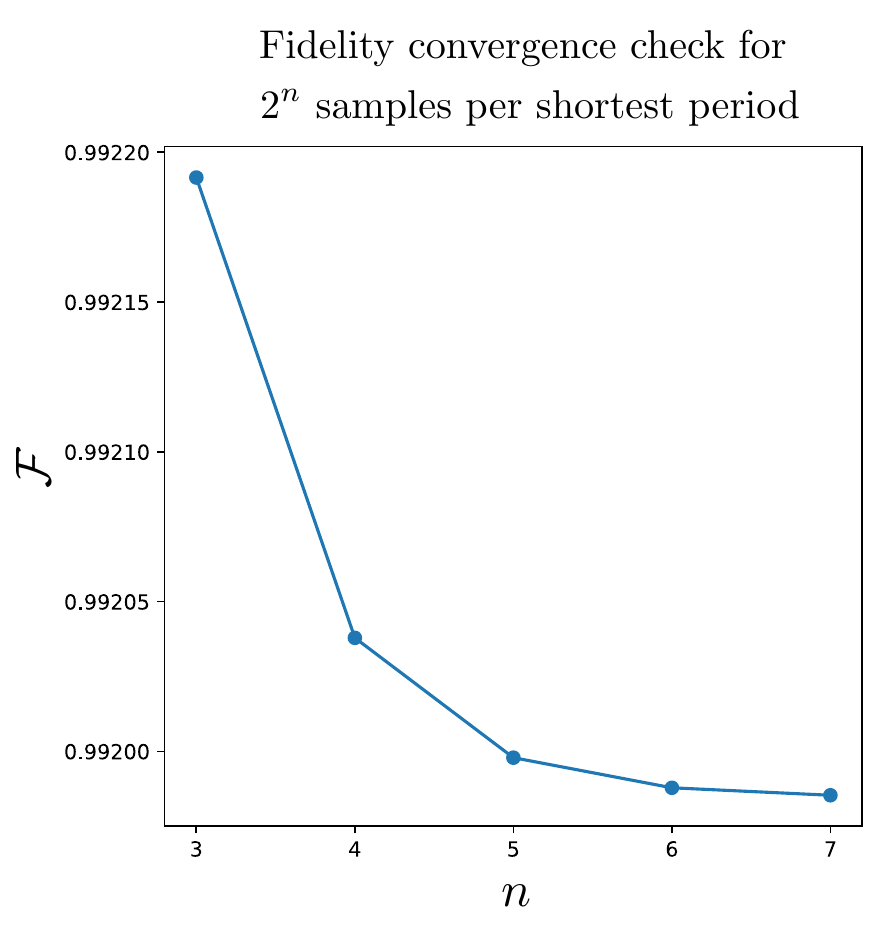}
    \caption{Fidelity at the optimal point of the spin qubit is plotted against different $2^n$ samples per shortest oscillation period. In our numerical calculations, the sampling number of $2^6 = 64$ is chosen to ensure numerical accuracy while rendering the computation time manageable. This gives an error in fidelity of less than $5 \times 10^{-6}$. Therefore, in the main text, we report fidelities $\mathcal{F}$ to 5 decimal places.}
    \label{fig:numerical convergence}
\end{figure}

To illustrate, for spin qubit, stage $\zeta = 3$ has the longest duration $\sqrt{2} \pi$ when with highest frequency $2\tilde \omega + 4\tilde \Omega$ is contained in two blue-sideband terms in the exact Hamiltonian, Eq.~\eqref{eq: spin exact Hamiltonian}. For the other stages, the assigned number of time steps is larger than 64 per shortest period of oscillation, thereby enhancing numerical convergence.The results of the numerical convergence at the optimal point are shown in Figure~\ref{fig:numerical convergence}, which justifies the choice of $2^6 = 64$ in Eq.~\eqref{eq: numerical samples number} that ensures numerical accuracy while rendering the computation time manageable.

For charge qubit, although the highest frequency in Hamiltonian~\eqref{eq: charge exact Hamiltonian} is infinity, the amplitude of high frequency terms is negligible, as discussed in Section~\ref{sec: fourier decomposition}. Therefore, we take the same sampling rate as for the spin qubit, without losing numerical convergence.

We utilize experimental parameters from Ref.~\cite{Mi2018}: photon frequency $\omega_r / 2\pi = 5.846 ~\text{GHz}$, the coupling strength $(g_s, g_c)/2\pi = (11~\text{MHz}, 40~\text{MHz})$, the qubit dephasing rate $(\gamma_s, \gamma_c)/2\pi = (5~\text{MHz}, 35~\text{MHz})$, and the photon loss rate $\kappa / 2\pi = 1.3~\text{MHz}$. The detuning $\Delta = 2\Omega$ is adjustable when the photon energy is fixed.

The normalized parameters are shown in Table~\ref{tab: normalized parameters}.
\begin{table}[!]
\begin{tabular}{|l|l|l|}
\hline
Normalized Parameters & Spin    & Charge  \\ \hline
$\tilde\omega_r$      & 531.455 & 292.30 \\ \hline
$\tilde\gamma$        & 0.455   & 1.750   \\ \hline
$\tilde\kappa$        & 0.118   & 0.065   \\ \hline
\end{tabular}
\caption{The normalized experimental parameters for both qubit systems from Ref.~\cite{Mi2018}.}
\label{tab: normalized parameters}
\end{table}

\section{Strong coupling gate time correction}
We eliminate the $\Zj$ terms in Eq.~\eqref{eq: spin exact Hamiltonian} by with an additional transformation~\cite{Ashhab2007, lu2012effects} $\hat U_5(t) = \exp[-\frac{i\Omega}{2\omega}\sin(2\omega t + 2\phi)\Zj]$, so the transformed spin qubit Hamiltonian $\hat H_{E, s}' = \hat U_5(t)^\dagger \left(\hat H_{E, s} - i\frac{\partial}{\partial t}\right) \hat U_5(t)$ is
\begin{align}
    \hat H_{E, s}' = &\Omega e^{2i\Omega t + iy\sin(2\omega t + 2\phi)} \sin(2\omega t + 2\phi) \pj + \Omega e^{-2i\Omega t - iy\sin(2\omega t + 2\phi)} \sin(2\omega t + 2\phi) \mj \nonumber \\
    &-g \a \Bigg( e^{-i\omega t + iy\sin(2\omega t + 2\phi)} \sin(\omega t + \phi) \pj + e^{-i(4\Omega + \omega) t - iy\sin(2\omega t + 2\phi)} \sin(\omega t + \phi) \mj \nonumber \\
    &+ e^{-i(2\Omega + \omega) t} \cos(\omega t + \phi)\Zj\Bigg) \\
    &-g \c \Bigg( e^{i(4\Omega + \omega) t + iy\sin(2\omega t + 2\phi)} \sin(\omega t + \phi) \pj + e^{i\omega t - iy\sin(2\omega t + 2\phi)} \sin(\omega t + \phi) \mj \nonumber\\
    &+ e^{i(2\Omega + \omega) t} \cos(\omega t + \phi) \Zj\Bigg)
    \nonumber
\end{align}
where $y = \Omega / 2\omega > 0$. Factorizing out the common function $g(t) = g \cos(y\sin(2\omega t + 2\phi))$ from sideband transition terms $\a\pj$ and $\c\mj$ and then dropping the rest of time-dependent terms, we have a new RWA sideband transition Hamiltonian
\begin{align}
    \HR^{(j)}(g(t), \phi) = \frac{ig(t)}{2} (e^{i \phi}\a\pj - e^{- i \phi} \c\mj),
\end{align}
where now, the effective coupling $g(t)$ is time-dependent.
To achieve the same CZ gate, equating
\begin{align}
    \int_{\tiz'}^{\tfz'} ds \cos(y\sin(2\omega s + 2\phi_s)) = \Delta \tau_s
\end{align}
we have the new gate time $\Delta \tau'_s$ for each step that can be solved using
\begin{align}
    J_0 (y) \Delta \tau'_s + \sum_{n = 1}^{\infty} \frac{J_{2n}(y)}{2\omega n} \sin(2n(2\omega t + 2\phi_s))\Big|^{\tiz' + \Delta \tau'}_{\tiz'} = \Delta \tau_s
\end{align}
where $J_m (z)$ are $m$-th order Bessel function. We have absorbed the real part of the evolution caused by $\Zj$ into the gate time.

However, numerical result shows that there are no improvement of fidelity with this strong-coupling correction of the gate time. The left qubit terms still act as an limit of high-fidelity region. One possible way of completely eliminate the qubit terms is to transform numerically according to
\begin{align}
    \hat U_5^\prime (t) = \mathcal{T}_{\leftarrow} \exp[-i\Omega \int_0^{t} dt \left( e^{2i\Omega t} \sin(2\omega t + 2\phi) \pj + e^{-2i\Omega t} \sin(2\omega t + 2\phi) \mj + \cos(2\omega t + 2\phi) \Zj\right)].
\end{align}
However, the cost of performing the transformation numerically is the loss of analytical insight.

\bibliography{Supp}